\documentclass[reqno, 11pt]{amsart}
\usepackage[mathscr]{eucal}

\usepackage{color}
 \usepackage[unicode,colorlinks,plainpages=false,hyperindex=true,bookmarksnumbered=true,bookmarksopen=true,pdfpagelabels]{hyperref}

\usepackage{amsfonts, amsmath, amsthm, amssymb, latexsym}
\newtheorem{theorem}{Theorem}[section]
\newtheorem{lemma}[theorem]{Lemma}
\newtheorem{prop}[theorem]{Proposition}
\newtheorem{cor}[theorem]{Corollary}
\theoremstyle{definition}
\newtheorem{defi}[theorem]{Definition}
\newtheorem{example}[theorem]{Example}
\theoremstyle{remark}
\newtheorem{remark}[theorem]{Remark}

\numberwithin{equation}{section}

\newcommand{\la}{\label}
\newcommand{\GL}{\mathrm{GL}}

\newcommand{\Res}{\mathrm{Res}}

\newcommand{\End}{\mathrm{End}}
\newcommand{\Mat}{\mathrm{Mat}}

\newcommand{\id}{\mathrm{id}}
\def\tr{\mathrm{tr}}
\def\Ind{\mathrm{Ind}}

\newcommand{\ov}{\overline}
\newcommand{\into}{\hookrightarrow}

\def\c{\mathbb{C}}
\def\R{\mathbb{R}}

\def\Z{\mathbb{Z}}

\def\N{\mathbb{N}}

\def\sl2{{\mathfrak{s}\mathfrak{l}}_2}
\def\vreg{V_{\rm{reg}}}
\def\VV{\widehat V}
\def\aalpha{\widehat{\alpha}}

\def\A{\mathcal{A}}
\def\S{\mathcal{S}}

\def\V{V_\R}

\def\D{\mathcal{D}}

\def\WW{W_a}
\def\Wh{\widehat W}
\def\Ah{A_\hbar}
\def\hb{\hbar}
\def\aalpha{\widetilde\alpha}
\def\HH{\widehat{\mathfrak H}}
\def\RR{R}

\def\Ra{R_a}

\newcommand\dpr[1]{\langle #1 \rangle}
\begin{document}
%
%
%
\title{Dunkl and Cherednik operators}
\author{Oleg Chalykh}
\address{School of Mathematics, University of Leeds, Leeds LS2 9JT, UK}
\email{o.chalykh@leeds.ac.uk}
%

%
\begin{abstract}
This survey article, written for the \emph{Encyclopedia of Mathematical Physics}, 2nd edition, is devoted to the remarkable family of operators introduced by Charles Dunkl and to their $q$-analogues discovered by Ivan Cherednik. The main focus is on the r\^ole of these operators in studying integrable many-body systems such as the Calogero--Moser and the Ruijsenaars systems. To put these constructions into a wider context, we indicate their relationship with the theory of the rational Cherednik algebras and double affine Hecke algebras. While we do not include proofs, references to the original research articles are provided, accompanied by brief historical comments.      
\end{abstract}

\maketitle

\section{Introduction}


Given a finite group $W$ acting linearly on $V=\c^n$, consider the action of $W$ on the algebra $\c(V)$ of meromorphic functions on $V$ (in particular, on the dual space $V^*$)  by $(w.f)(x)=f(w^{-1}.\,x)$ for $x\in V$, $w\in W$. The crossed product $\c(V)*W$ is formed by taking vector space $\c(V)\otimes\c W$
\footnote{Here and below tensor products are over $\c$, unless specified otherwise.} with the multiplication $(f\otimes w)(f'\otimes w')=f(w.f')\otimes ww'$. 
Let $\D(V)$ (respectively, $\D[V]$) denote the ring of differential operators on $V$ with meromorphic (respectively, polynomial) coefficients. It is generated by the directional derivatives $\partial_\xi$ with $\xi\in V$ and the operators of multiplication by $g\in \c(V)$ (respectively, by $g\in \c[V]$). We have a natural action of $W$ on $\D(V)$, hence the crossed product $\D(V)*W$, containing $\c(V)*W$ as a subalgebra. As an algebra, $\D(V)*W$ is generated by its two subalgebras, $1\otimes\c W$ and $\D(V)\otimes 1$ which can be identified with $\c W$ and $\D(V)$, respectively. Using these identifications, we replace $a\otimes w$ by $aw$, so each element of $\D(V)*W$ is written uniquely as $a=\sum_{w\in W}a_ww$ with $a_w\in\D(V)$. We may view such $a$ as operators on $\c(V)$, that is, elements of $\End_\c(\c(V))$, by composing the action of $W$ with that of $\D(V)$. 
 
Our typical setting below involves a real finite Coxeter group $W$ with its reflection representation $\V=\R^n$ equipped with a $W$-invariant Euclidean product $\dpr{\cdot, \cdot}$. Let $R\subset \V$ be the root system of $W$. For each root $\alpha\in R$, we have the orthogonal reflection $s_\alpha$  acting on $\V$ by the formula
\begin{equation*}
\la{refls}
s_\alpha.\,x = x-\dpr{\alpha^\vee, x}\alpha\,,\qquad \alpha^\vee:=
\frac{2\alpha}{\dpr{\alpha,\alpha}}\,.
\end{equation*}
The group $W$, generated by $s_\alpha$ with $\alpha\in R$, preserves the set $R$. Extending linearly over $\c$, we make $W$ act on $V=\c\otimes_\R\V$ equipped with a $W$-invariant symmetric bilinear form $\dpr{\cdot, \cdot}$.    

In this setting, Dunkl operators are a certain family of commuting elements of $\D(V)*W$. They depend on an additional parameter $k$, and for $k=0$ they are simply the directional derivatives $\partial_\xi$, $\xi\in V$. The Dunkl operators provide an interesting deformation of the partial derivatives, leading to a fruitful interplay with the theory of integrable systems, representation theory, special functions and harmonic analysis. 

If $W$, in addition, is a crystallographic reflection group, i.e., the Weyl group of a semisimple complex Lie algebra, then one can work with difference operators instead. Namely, we fix a $W$-invariant lattice $\Lambda\subset V$, pick $q=e^c$ with $c\ne 0$ and define $\D_q(V)$ as the subalgebra of $\End_\c(\c(V))$ generated by $\c(V)$ and $t(\lambda)=e^{c\partial_\lambda}$ with $\lambda\in \Lambda$. Each $t(\lambda)$ acts on $\c(V)$ by $t(\lambda). f\,(x)=f(x+c\lambda)$. We may, again, form the crossed product $\D_q(V)*W$ whose elements are written as $a=\sum_{w\in W}a_ww$ with $a_w\in\D_q(V)$, acting on $\c(V)$ in the obvious way. 

Cherednik operators are a non-trivial generalisation of the Dunkl operators: they are commuting elements of $\D_q(V)*W$ which can be viewed as a deformation of the operators $t(\lambda)$, $\lambda\in\Lambda$. Their theory is intrinsically linked to the theory of the double affine Hecke algebras (DAHAs) and Macdonald polynomials, while the Dunkl operators are similarly related to the rational version of DAHAs, the rational Cherednik algebras. There also exists an intermediate theory of degenerate DAHAs, related to the trigonometric Dunkl and rational Cherednik operators. 
Last but not least, both the Dunkl and Cherednik operators admit elliptic generalisations important for the study of the elliptic versions of the Calogero--Moser and Ruijsenaars systems. In particular, they can be used to construct quantum and classical Lax matrices for these systems.



\section{Dunkl operators for Coxeter groups and rational Calogero--Moser system}

\subsection{Dunkl operators} \label{2.1} Let $W$ be a finite Coxeter group with its reflection representation $V=\c^n$, equipped with a $W$-invariant complex Euclidean form $\dpr{\cdot, \cdot}$. Let $R\subset V$ be the root system of $W$. Each root $\alpha$ defines a linear function from $V$ to $\c$ by $x\mapsto \dpr{\alpha, x}$. This identifies $V$ with $V^*$, and the canonical pairing between $V, V^*$ with $\dpr{\cdot,\cdot}$.
Let us choose a $W$-invariant {\it multiplicity function} $k\,:\, R\to\c$, $\alpha\mapsto k_\alpha$. It is customary to partition $R$ into $R_+\sqcup -R_+$. Note that $k_\alpha=k_{-\alpha}$. If $W$ is irreducible,  $k$ takes on at most two different values.
The \emph{Dunkl operators} are the following elements of $\D(V)*W$:
\begin{equation}
\label{du} y_\xi :=
\hb\partial_\xi-\sum_{\alpha\in R_+}k_\alpha
\frac{\dpr{\alpha,\xi}}{\dpr{\alpha, x}}s_\alpha \,,\quad \xi\in V.
\end{equation}
Here $\hb\ne 0$ is a parameter\footnote{To follow the standard conventions of quantum mechanics, $\hb$ should be replaced $-\mathrm{i}\hbar$ and $k_\alpha$ should be purely imaginary.} which is only needed to handle the classical limit, otherwise one may assume $\hb=1$. 
The main two properties of the operators \eqref{du} 
are:

$(1)$\ {\it commutativity:}\ $\, y_{\xi}\,y_{\eta}=
y_{\eta}\,y_{\xi} \,$\quad $\forall\  \xi, \eta\in V$,

$(2)$\ {\it $W$-\,equivariance:}\ $\,w\,y_\xi w^{-1} =
y_{w.\xi}\,$ \quad$\forall\  w\in W$.

\noindent As a result, the assignment $\,\xi \mapsto y_\xi\,$
extends to a $W$-\,equivariant algebra map
\begin{equation}
\la{hom}
\c[V^*]\to \D(V)*W \,,\quad q \mapsto q(y) \ .
\end{equation}
The following version of the Dunkl operators is often used:
 \begin{equation}
\label{dut} T_\xi :=
\partial_\xi+\sum_{\alpha\in R_+}k_\alpha
\frac{\dpr{\alpha,\xi}}{\dpr{\alpha, x}}(1-s_\alpha) \,,\quad \xi\in V.
\end{equation}
The family $\{T_\xi\}$ is also commutative and $W$-\,equivariant, and has the advantage of preserving the polynomial subspace $\c[V]$ when acting on $\c(V)$.
  
\subsection{Rational Calogero--Moser system}
The operators \eqref{du} can be used to prove the integrability of the \emph{rational Calogero--Moser system}. The following calculation is important for that.
\begin{example}\label{cmd}
Let $\partial_i=\partial_{\xi_i}$ and $y_i=y_{\xi_i}$, where $\{\xi_i\,|\, i=1 \dots n\}$ is an orthonormal basis in $V$. For $\dpr{y, y}:=y_1^2+\dots +y_n^2$, a direct calculation shows that 
\begin{equation}\la{dcm}
\dpr{y, y}=\hb^2\Delta-\sum_{\alpha\in R_+} \frac{\dpr{\alpha, \alpha}}{\dpr{\alpha, x}^2}k_\alpha(k_\alpha-\hb s_\alpha)\,, \qquad \Delta=\sum_{i=1}^n \partial_{i}^2\,.
\end{equation} 
\end{example}
\noindent 
Introduce the linear map 
\begin{equation}\label{res}
\mathrm{Res}:\ \D(V)*W\,\to\, D(V)\,,\qquad \sum_{w\in W}a_ww\mapsto \sum_{w\in W}a_w\,.  
\end{equation}
Combining this map with \eqref{hom}, define
\begin{equation}\la{lp}
L_q=\Res\,q(y)
\,,\qquad 
q\in \c[V^*]^W\,.
\end{equation}
If $q$ is homogeneous, $L_q$ is a differential operator with the leading symbol $q(\hb\partial)$. The $W$-invariance of $q$ and the commutativity of the Dunkl operators imply that (1) each $L_q$ is $W$-invariant, (2) $L_qL_{q'}=L_{qq'}$ for any $q,q'$, (3) the family $\{L_q\,,\ q\in \c[V^*]^W\}$ is commutative. Hence, the following result.

\begin{theorem}\label{heckthm}
The map $q\mapsto L_q$ 
defines an algebra embedding $\c[V^*]^W\to\D(V)^W$. 
\end{theorem}
\noindent For any Coxeter group $W$, the algebra $\c[V^*]^W$ is a polynomial algebra on $n=\dim V$ generators, so this theorem produces $n$ algebraically independent differential operators $L_q\in\D(V)^W$. For $q=\dpr{\xi, \xi}$, we readily find from \eqref{dcm} that
\begin{equation}\label{cm}
L_{\dpr{\xi,\xi}}=\hb^2\Delta-\sum_{\alpha\in R_+} \frac{k_\alpha(k_\alpha-\hb)\dpr{\alpha, \alpha}}{\dpr{\alpha, x}^2}\,. 
\end{equation}
This is known as the Hamiltonian of the quantum rational Calogero--Moser system associated to the group $W$.
\begin{cor}
The quantum rational Calogero--Moser system \eqref{cm} is completely integrable, that is, it admits $n=\dim V$ pairwise commuting, algebraically independent Hamiltonians. 
\end{cor} 
The same construction, using $T_\xi$ instead of $y_\xi$, demonstrates the integrability of the Calogero--Moser Hamiltonian in ``radial form''
\begin{equation}\label{cmrad}
L=\Delta+\sum_{\alpha\in R_+} \frac{2k_\alpha}{\dpr{\alpha, x}}\partial_\alpha\,. 
\end{equation}

\begin{example}
Consider $W=S_n$ acting on $V=\c^n$ by permuting the basis vectors $e_1, \dots, e_n$; it is generated by permutations $s_{ij}$, $i\ne j$.  The ring $\c(V)$ is the ring of functions of $n$ variables $x_1,\dots, x_n$. The root system 
$
R=\{e_i-e_j\ |\ 1\le i\ne j \le n\}
$
is of type $A_{n-1}$, and the multiplicity function $k: R\to \c$ amounts to a single number $k\in \c$.  Taking $y_i=y_{e_i}$, we obtain 
$n$ commuting Dunkl operators
\begin{equation*}\label{dua}
y_i=\hb\frac{\partial}{\partial x_i}-k\sum_{j\ne i}\frac{1}{x_i-x_j}s_{ij}\,,\qquad i=1\dots n\,.
\end{equation*}
In this case the Calogero--Moser operator \eqref{cm} becomes
\begin{equation*}\label{cma}
L=\hb^2\Delta-\sum_{i<j} \frac{2k(k-\hb)}{(x_i-x_j)^2}\,. 
\end{equation*}
\end{example}

The Dunkl operators can also be used to construct the so-called {\it shift operators} that intertwine operators \eqref{lp} for different values of $k$. Let us write $L_{q}(k)$ to indicate the dependence of \eqref{lp} on the multiplicity function, $k$. Modify \eqref{res} by defining $\Res^-\left(\sum_{w\in W}a_ww\right)=\sum_{w\in W}\det(w)a_w$. 
\begin{theorem}\la{shift} Set $\hbar=1$ and
 define $S=S(k)\in \D(V)$ by $S=\Res^-\left(\prod_{\alpha\in R_+}y_{\alpha}\right)$. Then 
 \begin{equation}
  L_q(k+1)S(k)=S(k)L_q(k)\qquad\forall\ q\in\c[V^*]^W   
 \end{equation}
 where the multiplicity $k+1$ is defined by $(k+1)_\alpha=k_\alpha+1$ $\forall\ \alpha$.    
\end{theorem}

\subsection{Classical case}\label{clcase} 
The classical limit corresponds to taking $\hb\to 0$. More precisely, we view the Dunkl operators as elements of the algebra
\begin{equation*}
\Ah*W=\c(V)[\hat p_1, \dots, \hat p_n][[\hbar]]*W\,,
\end{equation*}
where the quantum momenta $\hat p_k=\hbar\partial_{\xi_k}$ satisfy the commutation relations $[\hat p_k, f]=\hbar\,\partial_{\xi_k}f$ for $f\in\c(V)$. We have an algebra isomorphism
\begin{equation*}
\eta_0:\ (\Ah*W)/\hbar(\Ah*W)\to A_0*W\,,\qquad f\mapsto f\,,\ \ \hat p_k\mapsto p_k\,,\ \ w\mapsto w\,,
\end{equation*}
where 
$A_0=\c(V)[p_1, \dots, p_n]=\c(V)\otimes \c[V^*]$ is the classical version of $\Ah$. 
Therefore, $\Ah$ (resp. $\Ah*W$) is a formal deformation of $A_0$ (resp. $A_0*W$). 
Note that $A_0$ is commutative, with the Poisson bracket defined by $\{\eta_0(a), \eta_0(b)\}=\eta_0(\hbar^{-1}[a,b])$ for $a,b\in \Ah$. For any $a\in\Ah*W$, we call $\eta_0(a)$ the \emph{classical limit} of $a$. Thus, the classical limit of \eqref{du} is the {\it classical Dunkl operator}
\begin{equation*}
y_{\xi, c}=p_\xi-\sum_{\alpha\in R_+}k_\alpha
\frac{\dpr{\alpha,\xi}}{\dpr{\alpha, x}} s_\alpha\,.
\end{equation*}
Here $p_\xi$ is the classical momentum in direction $\xi$. The operators $y_{\xi, c}$ are commuting elements of $A_0*W$, and we have a classical variant of \eqref{hom}:
\begin{equation}\label{homc}
\c[V^*] \to A_0*W \ ,\quad q \mapsto q(y_{c})\,.
\end{equation} 
\begin{theorem}\label{heckthmc}
The map $q\mapsto L_{q, c}=\Res\,q(y_c)$ 
defines an algebra embedding $\c[V^*]^W\to A_0^W$, whose image is a Poisson-commutative subalgebra $\mathcal A_c\subset A_0^W$. 
\end{theorem}
As a corollary, this proves integrability (in Liouville sense) of the classical rational Calogero--Moser system described by
\begin{equation}\label{cmcl}
L_{\dpr{\xi,\xi}, c}=\dpr{p, p}-\sum_{\alpha\in R_+} \frac{k_\alpha^2\dpr{\alpha, \alpha}}{\dpr{\alpha, x}^2}\,. 
\end{equation}

\subsection{Rational Cherednik algebras}
\la{S2}
These constructions can be put into a wider context. Given $V$, $W$ and the multiplicity function $k$ as above, we define the {\it rational Cherednik algebra}
$H_k=H_{\hb, k}$ as the subalgebra of $\D(V) *W$ generated by  $ \c W$, $\c[V]$,
and the image of $\c[V^*]$ under \eqref{hom}. Abstractly, $H_k$ can be described by generators $\xi\in V$, $\eta\in V^*$, $w\in W$ and relations
\begin{eqnarray}\la{chrel}
&&[\xi,\,\xi']=0\ ,\quad   [\eta,\,\eta'] =0 \ ,\quad
w\xi w^{-1}=w.\,\xi\,,\quad w\eta w^{-1}=w.\,\eta\ ,\nonumber \\*[1ex]
&& [\xi, \eta]=\hb\dpr{\xi, \eta}+
\sum_{\alpha\in R_+}k_\alpha \dpr{\alpha, \xi} \dpr{\alpha^\vee, \eta}s_\alpha \,.
\end{eqnarray}
The above realization of $\, H_k$ inside $\D(V)*W$, which sends 
\begin{equation}\label{chrep}
\xi\mapsto y_\xi,\quad \eta\mapsto \eta, \quad w\mapsto w\,,
\end{equation}
is referred to as the {\it Dunkl representation}. (Another representation of $H_k$ can be defined by sending $\xi$ to $T_\xi$.) The family $ \{H_k\} $ is a deformation (in fact, 
universal deformation) of $\, H_0 = \D[V]*W \,$. 
In particular, $H_k$ has the PBW property: a vector space isomorphism $H_k\stackrel{\sim}{\to}  \c[V]\otimes \c W \otimes \c[V^*]$. 
Recall the identification $*\,:\, V\to V^*$ defined by $\alpha^*=\dpr{\alpha,-}$, whose inverse is also denoted by $*$.
The symmetric nature of \eqref{chrel} implies that the mapping
\begin{equation}\label{ft}
V\ni\xi\mapsto \xi^*\,,\quad V^*\ni \eta\mapsto -\eta^*\,,\quad w\mapsto w  
\end{equation}
extends to an automorphism of $H_k$, which can be viewed as an (abstract) Fourier transform.

The {\it spherical subalgebra} of $ H_k
$ is defined as $e\, H_k \,e \,$, where $\, e =
|W|^{-1} \sum_{w \in W } w \,$ is the symmetrising idempotent. Restricting \eqref{res} onto the spherical subalgebra, one obtains an algebra map
 \begin{equation}
\la{HC}
\Res:\,e H_k e \into \D(V)^W \,.
\end{equation}
The family of algebras $B_k:=\Res(eH_ke)\subset \D(V)^W$ represent the universal deformation of $B_0=\D[V]^W$. The algebra $B_k$ has two commutative subalgebras: $\c[V]^W$ and the algebra of operators \eqref{lp}; the automorphism \eqref{ft} acts on $B_k$, exchanging them. 

The algebras $H_k$ and $B_k$ have rich representation theory. The important {\it category $\mathcal O$} is formed by the $H_k$-modules on which each $\xi\in V$ acts locally nilpotently. The structure of this category is determined by the {\it standard
modules}, which play a r\^ole similar to Verma modules in Lie theory.
To define such modules we fix an irreducible representation $ \tau $ of $W$
and extend the $W$-structure on $ \tau $ to a $\,\c[V^*] * W$-module
structure by letting $\,\xi \in V \,$ act trivially.
The {\it standard $H_k$-module of type} $ \tau $ is then given by
\begin{equation}\la{sm}
M(\tau) := \mathrm{Ind}^{H_k}_{\c[V^*]*W}\,\tau\, =\,
H_k\underset{\c[V^*]*W}{\otimes} \tau\,.
\end{equation}
The PBW property of $H_k$ implies that $\, M(\tau) \cong \c[V]\otimes \tau \,$
as a $\c[V]$-module. An important example is the trivial representation $\tau$, in which case $\, M(\mathrm{triv}) \cong \c[V]\,$ defines the so-called {\it polynomial representation} of $H_k$. One can check that the action of any $\xi\in V$ on this space is given by the Dunkl operator \eqref{dut}.

\subsection{KZ connection} The Dunkl operators and Cherednik algebra can be used to construct a family of flat connections known as Knizhnik--Zamolodchikov (KZ) connections. Let $$\delta(x)=\prod_{\alpha\in R_+} \dpr{\alpha, x}\quad\text{and}\quad \vreg=\{x\in V\,|\, \delta(x)\ne 0\}\,.$$ Correspondingly, we have the algebras $\c[\vreg]$ and $\D[\vreg]$ of regular functions and differential operators on $\vreg$. 
Write $H_k[\delta^{-1}] \,$ for the 
localization of $H_k$ on (powers of) $\delta$. From the Dunkl representation \eqref{chrep} it follows easily  
that 
one may identify $H_k[\delta^{-1}] \, \cong \D[\vreg]*W$. Applying the same localisation to the standard module $M=M(\tau)$ \eqref{sm} produces the space $M[\delta^{-1}]\cong \c[\vreg]\otimes\tau$, with an action of  $\D[\vreg]*W$. This allows us to view $M[\delta^{-1}]$ as the space of sections of 
a trivial vector bundle on $\vreg$ of rank $\dim\tau$, and a $\D[\vreg]$-module. A direct calculation shows that the $\D$-module structure on $M[\delta^{-1}]$ is
described by
\begin{equation}\la{conn}
\hb\partial_\xi(f\otimes v) = \hb\partial_\xi(f)\otimes v + \sum_{\alpha\in R_+}
k_\alpha \frac{\dpr{\alpha,\xi}}{\dpr{\alpha, x}}f\otimes
s_\alpha v\ ,
\end{equation}
which defines a regular flat connection
on $ \c[\vreg] \otimes \tau $. It is
called a {\it KZ connection} with values in $ \tau \,$: its
horizontal sections $\,\varphi: \vreg \to \tau \,$ satisfy the
{\it KZ equations}
\begin{equation}\label{kz}
\hb\partial_\xi \varphi + 
\sum_{\alpha\in R_+}
k_\alpha \frac{\dpr{\alpha,\xi}}{\dpr{\alpha, x}}s_\alpha(\varphi)=0
\ ,\quad \forall\,\xi\in V\ .
\end{equation}
An important
difference compared to \eqref{du} is that in \eqref{kz}, the reflections $s_\alpha$  act on the {\it values} of
the functions involved, while in \eqref{du} they act on their arguments.

\subsection{Dunkl transform}
 
In this section we assume that $k_\alpha\ge 0$ for all $\alpha\in R$. 
Given $y\in V$, consider the joint eigenvalue problem for the Dunkl operators \eqref{dut}:
\begin{equation}\label{ek}
T_\xi f= \dpr{\xi, y}f\,,\quad \forall\ \xi\in \V\,.
\end{equation} 

\begin{theorem} The system \eqref{ek} has a unique solution $f: x\mapsto E_k(x ,y)$ which is real-analytic on $\V$ and satisfies $f(0) = 1$. Moreover, the mapping $(x, y) \mapsto E_k(x, y)$ extends to a holomorphic function on $V\times V$. 
\end{theorem}
\noindent The function $E_k$ is called the {\it Dunkl kernel}.  It has the following properties:

$(1)$ $E_k(x,y)=E_k(y,x)$,

$(2)$ $E_k(\lambda x, y)=E_k(x, \lambda y)$ for any $\lambda\in \c$,

$(3)$ $E_k(w.x, w.y)=E_k(x, y)$ for any $w\in W$,

$(4)$ $\overline{E_k(x,y)}=E_k(\overline{x}, \overline{y})$,

$(5)$ $E_0(x,y)=\exp(\dpr{x, y})$. 

\noindent Let 
\begin{equation}\label{ck}
w_k(x)= \prod_{\alpha\in R_+} |\dpr{\alpha, x}|^{2k_\alpha}\,,\qquad c_k=\int_{\V} e^{-|x|^2/2}w_k(x)\,\mathrm{d}x\,.
\end{equation}
The {\it Dunkl transform} associated with $W$ and the multiplicity function $k$ is defined by
\begin{equation}
\widehat{f}(\xi)=c_k^{-1}\int_{\V} f(x) E_k(-i\xi, x)w_k(x)\,\mathrm{d}x\qquad (\xi\in\V)\,.
\end{equation}
The {\it inverse transform} is defined by $f^\vee(\xi)=\widehat{f}(-\xi)$. 
\begin{theorem}\label{transform} $(1)$ The Dunkl transform is a homeomorphism of the Schwartz space $\mathcal{S}(\V)$.

$(2)$ (Plancherel theorem) The Dunkl transform has a unique extension to an isometric isomorphism of $L^2(\V, w_k)$.  

$(3)$ ($L^1$-inversion) If both $f$ and $\widehat f$ belong to $L^1(\V, w_k)$, then $f=(\widehat f)^\vee$ almost everywhere.

\end{theorem}

\subsection{Generalisation for complex reflection groups} For this section only, $W$ is 
a finite subgroup of $ \GL(V) $ generated by complex
reflections. An element $ s \in W$, $s\ne\id$, is
a {\it complex reflection} if it
acts as
identity on some hyperplane $ H_s $ in $ V $.
Let $\S$ denote the set of complex reflections $s$ in $W$; the group $ W $ acts on $\S$ by conjugation.
Write $ \A $ for the set of reflection hyperplanes  $\,\{H_s\}\,$ for reflections in $W$. 
If $ H \in \A $, the (pointwise) stabilizer of $H$ in $W$
is a finite cyclic subgroup $ W_H \subseteq W $. 
For each $H\in\A$ we fix a covector $ \alpha_H \in V^*$  with $\ker \alpha_H=H$. 

Given a  $W$-invariant multiplicity function $k\,:\, \S\to \c$, $s\mapsto k_s$, we define the {\it Dunkl operators}
as the following elements of $\D(V)* W \,$:
\begin{equation}
\label{duc} y_\xi :=
\hb\partial_\xi-\sum_{H\in \A}
\frac{\alpha_H(\xi)}{\alpha_H}\sum_{s\in W_H\setminus\{\id\}}k_ss\ , \quad \xi \in V\ .
\end{equation}
These operators have the same basic properties as in the Coxeter case, namely, commutativity and $W$-equivariance. Similarly, the assignment $\,\xi \mapsto y_\xi\,$
extends to an injective algebra homomorphism
\begin{equation}
\la{homcc}
\c[V^*] \hookrightarrow \D(V)* W \ ,\quad q \mapsto q(y) \ .
\end{equation}
%
With these definitions, Theorem \ref{heckthm} remains valid. 
The algebra $\c[V^*]^W$ is, again, a polynomial algebra on $n=\dim V$ generators, so this gives $n$ algebraically independent differential operators \eqref{lp}. If $W$ is non-Coxeter (and irreducible), all elements in $\c[V^*]^W$ have degrees $>2$ and, in general, the quantum Hamiltonians $L_q$ are too complicated to allow any explicit expression. 

One also introduces the rational Cherednik algebra, category $\mathcal O$ and KZ connection in the same  way as it was done in the Coxeter case. 

\subsection{Historical comments} The Dunkl operators for Coxeter groups were introduced by Dunkl in \cite{D1}, where their commutativity and the formula \eqref{dcm} was established. Theorems \ref{heckthm}, \ref{shift} are due to Heckman \cite{He}. Similar constructions in the case $W=S_n$ can be found in the physics literature \cite{BHV, P}, including the case of the Calogero--Moser system with the oscillator term (cf. \cite{FLP, FeH} for some recent developments). Integrability of Calogero--Moser systems in some cases was known before, see \cite{OP}, but no full (and uniform) proof was available prior to \cite{He}. The existence of shift operators (in trigonometric case) was first proved by Opdam in \cite{HO4} but Heckman's proof based on Dunkl operators offered a drastic simplification.    

Rational Cherednik algebras were introduced and studied by Etingof and Ginzburg in \cite{EG} as a rational counterpart of Cherednik's theory of DAHAs. 
Classical limit of the Dunkl operators and the algebra $H_k$ are discussed in \cite{EG, Et}. 
The category $\mathcal O$ for $H_k$ first appeared in \cite{DO}, followed by \cite{GGOR, BEG}.  The link between Dunkl operators and KZ equations in the trigonometric setting goes back to Cherednik \cite{C1}; in the present setting it appeared in \cite{GGOR, BEG}. There is a correspondence between solutions to the KZ equations and the eigenfunctions of the operators \eqref{lp}, called {\it Matsuo--Cherednik correspondence} \cite{Ma, C94}, see also \cite{FV} for an interpretation of the shift operators from that viewpoint. Dunkl operators and Cherednik algebras can be used to study deformed Calogero--Moser systems and generalisations, see \cite{SV09, Fe, FeS, BC1}.

The Dunkl kernel and Dunkl transform are due to Dunkl \cite{D2}, see also Opdam's work \cite{O1} for the in-depth study of the eigenvalue problem  \eqref{ek} and of the joint eigenfunctions of the operators $L_q$. The constant $c_k$ in \eqref{ck} is known explicitly, see \cite{E1} and references therein. Theorem \ref{transform} is due to Dunkl and de Jeu \cite{D3, dJ}, see \cite{R} for further results and references. 

For complex reflection groups, Dunkl operators and operators $L_q$ were constructed, and their commutativity proved, by Dunkl and Opdam in \cite{DO}. For further results, see \cite{O2, BC}.

\section{Affine Hecke algebras and Cherednik operators}

\subsection{Affine Weyl groups and Hecke algebras}\la{3.1}

Let $R\subset \V$ be a reduced, irreducible crystallographic root system,
and $W$ be the Weyl group of $R$, generated by the orthogonal reflections $s_\alpha$, $\alpha\in R$. As before, we extend the action of $W$ onto $V$ by $\c$-linearity. 
We write $R^\vee=\{\alpha^\vee\}$ for the dual system formed by the coroots $\alpha^\vee=2\alpha/\dpr{\alpha, \alpha}$. 
Let $a_1, \dots, a_n$ be a fixed basis of simple roots in $R$, associated with a decomposition $R=R_+\sqcup R_-$. We have the coroot and coweight lattices: $Q^\vee=\sum_{i=1}^n \Z a_i^\vee$ and $P^\vee=\sum_{i=1}^n \Z b_i$, where the fundamental coweights $b_i\in V$ are defined by $\dpr{a_i, b_j}=\delta_{ij}$. We write $Q^\vee_+=\sum_{i=1}^n \Z_{\ge 0} a_i^\vee$ and $P^\vee_+:=\sum_{i=1}^n\Z_{\ge 0} \,b_i$ for the cones of {\it positive} coroots and {\it dominant} coweights, respectively.

The \emph{affine Weyl group} is defined as $\WW=W\ltimes t(Q^\vee)$, where $t(Q^\vee)$ denotes the group of translations $t(\lambda)$, $\lambda\in Q^\vee$ acting on $V$ by 
$t(\lambda).x=x-c\lambda$, where $c$ is a fixed parameter. The \emph{extended} affine Weyl group is $\Wh=W\ltimes t(P^\vee)$. The group $\Wh$ acts naturally on functions on $V$. In particular, a translation $t(\lambda)$, $\lambda\in P^\vee$ acts by
$t(\lambda).f(x)=f(x+c\lambda)$, i.e., $t(\lambda)=e^{c\partial_\lambda}$.

Let $\VV$ denote the space of affine-linear complex-valued functions on $V$. We identify $\VV$ with $V\oplus\c\delta$, where vectors in $V$ are considered as linear functionals on $V$ via the scalar product $\dpr{\cdot , \cdot}$ and where $\delta\equiv c$ on $V$
(so $e^\delta=q:=e^c$).
Let
\begin{equation}\la{rrelq}
\Ra=\{\aalpha=\alpha+m\delta\,, m\in\Z\,, \alpha\in R\}\subset \VV
\end{equation} 
be the affine root system associated with $R$. The action of $\Wh$ on $\VV\subset \c(V)$ permutes affine roots. For any $\aalpha=\alpha+m\delta$ we have the orthogonal reflection with respect to the hyperplane $\aalpha(x)=0$ in $V$,
\begin{equation*}
s_{\aalpha}.\, x=x-\aalpha(x)\alpha^\vee\,,\quad x\in V\,.
\end{equation*}
We extend the set of simple roots $a_i$ to a basis in $\Ra$ by adding $a_0=\delta-\varphi$, where $\varphi$ is the highest root in $R_+$. 
The simple reflections $s_{i}=s_{a_i}$, $i=0,\dots, n$ generate the group $\WW$, and the {\it length} $l(w)$ of $w\in\WW$ is defined as the length $l$ of a reduced (i.e., shortest) decomposition 
\begin{equation}\la{red}
w=s_{i_1}\dots s_{i_l}\,,\quad\text{with}\ 0\le i_k\le n\,.
\end{equation}
Let $\Omega$ be the subgroup of the elements $\omega\in\Wh$ which map the basis $a_0, \dots, a_n$ to itself. It is known that $\Omega$ is an abelian group, isomorphic to $P^\vee/Q^\vee$, and the extended affine Weyl group is isomorphic to $\WW\rtimes\Omega$. Each $\tilde w\in\Wh$ admits a unique presentation as $\tilde w=w \omega$ with $w\in\WW$ and $\omega\in\Omega$. We use this to extend the notion of the length from $\WW$ to $\Wh$ by setting $l(w\omega)=l(w)$, so $l(\omega)=0$ for all $\omega\in\Omega$. 

The \emph{braid group} $\mathfrak B$ of $\Wh$ is the group with generators $T_w$, $w\in\Wh$, and relations 
\begin{equation}\label{brad}
T_vT_w=T_{vw}\quad\text{if}\quad l(v)+l(w)=l(vw)\,.
\end{equation} 
Write $T_i:=T_{s_i}$ for $i=0, \dots, n$. Then for any reduced decomposition $w=s_{i_1}\dots s_{i_l}\omega$ we have
$T_w=T_{i_1}\dots T_{i_l}T_\omega$ . It follows that $\mathfrak B$ is generated by $T_i$, $i=0, \dots, n$ and $T_\omega$, $\omega\in\Omega$, subject to the following relations: 
\begin{align}\la{braid}
&T_{i}T_{j}\dots=T_{j}T_{i}\dots \qquad\text{for $i\ne j$, with $m_{ij}$ factors on either side}\,,\\\la{braid1}
&T_\omega T_{\omega'}=T_{\omega \omega'}\quad\text{for}\ \omega, \omega'\in\Omega\,,\\\la{braid2}
&T_\omega T_iT_\omega^{-1}=T_j\qquad\text{if}\ \omega s_i\omega^{-1}=s_j\,.
\end{align}
Here $m_{ij}=2, 3, 4, 6$ is the order of $s_is_j\in \WW$.

The braid group $\mathfrak B$ contains an important abelian subgroup $\{Y^\lambda\,|\,\lambda\in P^\vee\}$. 
Namely, if $\lambda, \mu\in P^\vee$ are dominant, then from \eqref{brad} $T_{t(\lambda)} T_{t(\mu)}=T_{t(\mu)}T_{t(\lambda)}=T_{t(\lambda+\mu)}$. Hence, $Y^\lambda:=T_{t(\lambda)}$ with $\lambda\in P_+^\vee$ form an abelian semigroup, and we extend it by setting  $Y^\lambda=Y^\mu(Y^\nu)^{-1}$ whenever $\lambda=\mu-\nu$ with $\mu, \nu\in P^\vee_+$. 

Choose nonzero parameters $\tau_i$, $i=0, \dots, n$ such that $\tau_i=\tau_j$ if $s_i$ and $s_j$ are conjugated in $\Wh$. We use $\tau$ to denote the set of parameters. Note that $\tau_i$'s take on at most  two different values.  The (extended) \emph{affine Hecke algebra} $\HH$ is the quotient of the group algebra $\c\mathfrak B$ by relations 
\begin{equation}\la{hecke}  
(T_i-\tau_i)(T_i+\tau_i^{-1})=0\,,\qquad i=0,\dots n\,.
\end{equation}
For $\tau_i=1$ we have $\HH\cong \c\Wh$. The subalgebra $\mathfrak H\subset \HH$ generated by $T_1, \dots, T_n$ is known as the (finite) {\it Hecke algebra} of $W$.
The elements $T_w$, $w\in\Wh$ ($w\in W$, respectively) form a $\c$-basis of $\HH$ ($\mathfrak H$, respectively). An alternative basis for $\HH$ is given by the elements $T_w Y^\lambda$ with $w\in W$, $\lambda\in P^\vee$.

\subsection{Basic representation and Cherednik operators}\la{bas} Setting $q=e^c$, let $\D_q(V)\subset \End_\c(\c(V))$ be the algebra of difference operators for the lattice $\Lambda=P^\vee$. We view the group algebra $\c\Wh$ as a subalgebra of $\D_q(V)*W$, by sending 
\begin{equation}\la{hatin}
\Wh\ni wt(\lambda)\mapsto we^{c\partial_\lambda}\qquad (w\in W, \ \lambda\in P^\vee)\,. \end{equation}
The algebra $\widehat{\mathfrak H}$ can also be realized as a subalgebra of $\D_q(V)*W$. This is called the {\it basic representation} of $\HH$.   To describe it, extend the set of parameters $\tau_i$ to $\tau_\alpha$, $\alpha\in\Ra$ so that $\tau_\alpha=\tau_{w(\alpha)}$ for $w\in\Wh$, and introduce $\mathbf{c}_\alpha\in \c(V)$ as follows:
\begin{equation}\la{cal}
\mathbf{c}_\alpha=\frac{\tau_\alpha^{-1}-\tau_\alpha e^{\alpha}}{1-e^{\alpha}}\,,\qquad\alpha\in\Ra\,.
\end{equation}
\begin{theorem}\la{bre}
\la{beta} The extended affine Hecke algebra admits a faithful representation $\beta: \widehat{\mathfrak H} \to \D_q(V)*W$ such that   
\begin{align}\la{br}
&\beta:\,T_i\mapsto \tau_i+\mathbf{c}_i(s_{i}-1)\,, \qquad \mathbf{c}_i=\mathbf{c}_{a_i} \qquad (i=0, \dots, n)\,,\\\la{br1}
&\beta:\,T_\omega\mapsto \omega\quad\text{for}\quad \omega\in\Omega\,.
\end{align}
\end{theorem}
The \emph{Cherednik operators} are, by definition, the images of $Y^\lambda$ under $\beta$. They form a commutative family of difference-reflection operators, and should be viewed as $q$-analogues of the Dunkl operators. In comparison, they are rather complicated. To find $Y^\lambda$ for $\lambda\in P_+^\vee$, for example,  one first finds a reduced decomposition $t(\lambda)=s_{i_1}\dots s_{i_l}\omega$ and then writes the product $Y^\lambda=T_{i_1}\dots T_{i_l}T_\omega$ in the basic representation.

From now on, we identify $\widehat{\mathfrak H}$ with its image under $\beta$, in particular, making no distinction between $Y^\lambda$ and the Cherednik operators. It is sometimes more convenient to write $Y^\lambda$ in terms of the following elements $\RR(\alpha)$ (``$R$-matrices''):
\begin{equation}\la{rdef}
\RR(\alpha)=\tau_{\alpha} s_{\alpha} +\mathbf{c}_{\alpha}(1-s_{\alpha})\,,\quad \alpha\in \Ra\,. 
\end{equation} 
These elements satisfy the property
\begin{equation*}
w\RR(\alpha)w^{-1}=\RR(w.\alpha)\,,\qquad\text{for any}\  w\in\WW\,. 
\end{equation*}
Using this and the fact that $T_i=\RR(a_i)s_i$ for $i=0, \dots, n$, it is straightforward to rewrite $Y^\lambda$ in terms of $\RR(\alpha)$ instead of $T_i$. 

The commutative subalgebra generated by the Cherednik operators will be denoted as $\c[Y]$, so elements $f(Y)\in\c[Y]$ are arbitrary linear combinations of $Y^\lambda$, $\lambda\in P^\vee$. Inside $\c[Y]$ we have the subalgebra $\c[Y]^W$, spanned by the orbitsums $f=\sum_{\mu\in W\lambda}Y^{\mu}$. 

\subsection{$\GL_n$-case}\la{aq} For the root system $R$ of type $A_{n-1}$ the above theory has another variant, referred to as the $\GL_n$-case. 
We start with $V=\c^n$, with the orthonormal basis $\epsilon_1, \dots, \epsilon_n$ and the associated coordinates $x_1, \dots, x_n$. The roots in $R$ are $\alpha=\epsilon_i-\epsilon_j$ with $i\ne j$, and the group $W={S}_n$ acts on $V$ by permuting the basis vectors. Instead of $P^\vee(R)$ we choose the lattice $\Lambda$ to be
$\Z^n=\sum_{i=1}^n \Z\epsilon_i$. There is only one parameter $\tau$ so  $\tau_i=\tau$. The algebra of difference operators ${\D}_q=\c(x)\ltimes t(\Z^n)$, associated with $\Lambda$, is generated by $\c(x)$ and $t(\epsilon_k)=e^{c\partial_k}$ ($1\le k\le n$). 
The simple roots are $a_i=\epsilon_i-\epsilon_{i+1}$ ($1\le i<n$), with simple reflections $s_{a_i}=s_{i,i+1}$. 

The extended affine Weyl group $\WW=S_n\ltimes \Z^n$ is generated by $s_{i, i+1}$ and $\omega$ acting on $f\in\c(x)$ by
\begin{equation}
(\omega.f)(x)=f(x_2, \dots, x_n, x_1-c)\,.
\end{equation}
The subgroup $\Omega$ (of elements of length zero) is generated by $\omega$. The algebra $\HH$ is generated by $T_1, \dots, T_{n-1}$ and $T_\omega$, subject to relations \eqref{braid1}, \eqref{hecke}, the relations that $T_{\omega}^n$ is central in $\HH$ and that $T_\omega T_i= T_{i+1}T_\omega$ for $1\le i<n$. The basic representation $\beta\,:\, \HH\to \D_q*S_n$ is defined by
\begin{equation}
\beta(T_\omega)=\omega\,,\quad \beta(T_i)=\tau+\mathbf{c}_{i, i+1}(s_{i,i+1}-1)\,,\qquad \mathbf{c}_{ij}=\frac{\tau e^{x_i}-\tau^{-1}e^{x_{j}}}{e^{x_i}-e^{x_{j}}}\,.
\end{equation}
The commuting Cherednik operators $Y_i=Y^{\epsilon_i}$ are 
\begin{equation}\la{yit}
Y_i=T_{i}T_{i+1}\dots T_{n-1}\,T_\omega\, T_{1}^{-1}\dots T_{i-1}^{-1}\quad (i=1, \dots, n)\,.
\end{equation}  
Using the elements
\begin{equation}\la{rij}
\RR_{ij}=\tau s_{ij}+\mathbf{c}_{ij}(1-s_{ij})\,,
\end{equation}
these can be rewritten as
\begin{equation}\la{yi}
Y_i=\RR_{i, i+1}\RR_{i, i+2}\dots \RR_{i, n}\,t(\epsilon_i)\,\RR_{1i}^{-1}\dots \RR_{i-1, i}^{-1}\quad (i=1, \dots, n)\,.
\end{equation}

\subsection{$C^\vee C_n$ case}\la{cc} Another special case is related to the non-reduced affine root system of type $C^\vee C_n$. 
Let $V=\c^n$ with the standard orthonormal basis $\{\epsilon_i\}$ and the associated coordinates $\{x_i\}$. Let  $R$ be the root system of type $C_n$,
\begin{equation}\la{cn}
R=\{\pm 2\epsilon_i\,|\,1\le i\le n\}\cup \{\pm \epsilon_i\pm \epsilon_j\,|\, 1\le i<j\le n\}\,.
\end{equation}
The Weyl group $W={S}_n\ltimes \{\pm 1\}^n$ of $R$ consists of the transformations that permute the basis vectors $\epsilon_i$ and change their signs arbitrarily. As in \ref{3.1}, we write $\VV=V\oplus\c\delta$ for the space of affine-linear functions on $V$, with $\delta\equiv c$. 
Let $\Ra$ be the affine root system associated with $R$ \eqref{rrelq}. As a basis of simple roots, we choose
\begin{equation*}
a_0=\delta-2\epsilon_1\,,\qquad a_i=\epsilon_i-\epsilon_{i+1}\quad (i=1,\dots, n-1)\,,\qquad \alpha_n=2\epsilon_n\,.
\end{equation*}
For this case, $\Wh=\WW=W\ltimes \Z^n$ is generated by $s_{i}=s_{a_i}$, $i=0,\dots, n$. Their action of the generators in coordinates is by
\begin{align}\la{wwact}
s_{0}\,(x_1,\dots, x_n)&=(c-x_1, x_2, \dots, x_n)\,,\nonumber\\
s_{i}\,(x_1,\dots, x_n)&=(x_1,\dots, x_{i-1}, x_{i+1}, x_i, \dots, x_n)\quad (i=1, \dots, n-1)\,,\\ 
s_{n}(x_1, \dots, x_n)&=(x_1, \dots, x_{n-1}, -x_n)\,.\nonumber
\end{align}
As in the $\GL_n$-case, we consider the algebra $\D_q$ of difference operators in $n$ variables, and view $\c\Wh$ as a subalgebra in ${\D}_q * W$.   

The algebra $\HH$ is generated by $T_0, \dots, T_n$ subject to the relations
\begin{align}
&T_iT_{i+1}T_iT_{i+1}=T_{i+1}T_iT_{i+1}T_i\quad (i=0, i=n-1)\,,\la{b1}\\
&T_iT_{i+1}T_i=T_{i+1}T_iT_{i+1}\quad (i=1, \dots, n-2)\,,\la{b2}\\
&T_iT_{j}=T_jT_{i}\,,\quad |i-j|\ge 2\,,\la{b3}\\
&(T_i-\tau_i)(T_i+\tau_i^{-1})=0\quad(i=0,\dots, n)\,,
\end{align}
where $\tau_i$ are deformation parameters. Here it is assumed that $\tau_1=\dots =\tau_{n-1}=\tau$, so $\HH$ depends on three parameters: $\tau_0, \tau_n$ and $\tau$. 

The basic representation $\beta: \HH \to {\D}_q*W$ involves two additional parameters, $\tau_0^\vee, \tau_n^\vee$. Let us introduce parameters $\tau_{\alpha}$ and functions $\mathbf{c}_{\alpha}$ for $\alpha\in \Ra$ as follows:
\begin{align*}
\tau_{\alpha} &=\tau\,,\quad \mathbf{c}_{\alpha} =\frac{\tau^{-1}-\tau e^{\alpha}}{1-e^{\alpha}}\,\quad \text{for\ }\alpha=k\delta\pm\epsilon_i\pm\epsilon_j\quad(k\in\Z,\ i\ne j)\,, \\
\tau_{\alpha} &=\tau_0\,, \quad \mathbf{c}_{\alpha} =\tau_0^{-1}\frac{(1-\tau_0\tau_{0}^\vee e^{\alpha/2})(1+\tau_{0}(\tau_{0}^\vee)^{-1} e^{\alpha/2})}{(1-e^{\alpha})}\,\quad \text{for\ }\alpha=(2k+1)\delta\pm 2\epsilon_i\quad(k\in\Z)\,,\\
\tau_{\alpha} &=\tau_n\,,\quad \mathbf{c}_{\alpha} =\tau_n^{-1}\frac{(1-\tau_n\tau_{n}^\vee e^{\alpha/2})(1+\tau_{n}(\tau_{n}^\vee)^{-1} e^{\alpha/2})}{(1-e^{\alpha})}\,\quad \text{for\ }\alpha=2k\delta\pm 2\epsilon_i\quad(k\in\Z)\,.
\end{align*}
With this notation, we define $\beta$ on generators (and extend by multiplicativity) by
\begin{equation*}
\beta:\, T_i\mapsto \tau_i+\mathbf{c}_{a_i} (s_{i}-1)\,, \quad i=0, \dots, n\,.
\end{equation*}
This defines a subalgebra of ${\D}_q*W$, isomorphic to $\HH$ and depending on five parameters $\tau_0, \tau_0^\vee, \tau_n, \tau_n^\vee, \tau$.
The commutative subalgebra $\c[Y]$ is generated by the operators $Y_i^{\pm 1}$, where $Y_i=Y^{\epsilon_i}$ is
\begin{equation*}
Y_i=T_i\dots T_{n-1}T_nT_{n-1}\dots T_1T_0T_1^{-1}T_2^{-1}\dots T_{i-1}^{-1}\,,\quad i=1,\dots, n\,.
\end{equation*}

\subsection{Macdonald--Ruijsenaars operators}\la{mrh} Similarly to Dunkl operators, symmetric combinations of the Cherednik operators produce commuting operators. The map \eqref{res} becomes  
\begin{equation}\label{resq}
\mathrm{Res}:\ \D_q(V)*W\,\to\, D_q(V)\,,\qquad \sum_{w\in W}a_ww\mapsto \sum_{w\in W}a_w\,.  
\end{equation}
\begin{theorem}\label{qheckthm}
The map $f\mapsto L_f=\Res \,f(Y)$ defines an algebra embedding $\c[Y]^W\to\D_q(V)^W$. 
\end{theorem}
The commutative family $\{L_f\,,\, {f\in\c[Y]^W}\}$ defines a quantum completely integrable system on $V$. Being difference operators, the Hamiltonians $L_p$ depend exponentially on quantum momenta $\widehat p=-\mathrm{i}\hbar\partial$. 

\begin{example} In the notation of \ref{aq}, let $f_r=e_r(Y_1, \dots, Y_n)$ be the elementary symmetric combinations of $Y_i$, $1\le r\le n$. Then the operators $L_r=\Res \, f_r(Y)$ have the following form:
\begin{equation}
\label{ru} L_r=\sum_{\genfrac{}{}{0pt}{}{I\subset
\{1,\dots,n\}}{|I|=r}} \prod_{\genfrac{}{}{0pt}{}{i\in I} {j\notin
I}} 
\mathbf{c}_{ij}\,\prod_{i\in I} t(\epsilon_i)\,.
\end{equation}
Up to a gauge transformation, these are commuting Hamiltonians of the {\it trigonometric Ruijsenaars system}, a relativistic version of the Calogero--Moser system.
\end{example}

Standard generators of $\c[Y]^W$ are the orbitsums for the fundamental coweights $b=b_i$, so we denote
\begin{equation}
f_b =\sum_{\pi\in Wb} Y^\pi\,,\qquad L_b=\Res \,(f_b)\,.
\end{equation}
The operators $L_b$ are complicated in general. The following result gives explicit expressions for some of them, known as {\it Macdonald operators}. 

\begin{theorem}\label{mo}  (i) Let $b\in P^\vee_+$ be minuscule, so that $\dpr{\alpha, b}$ is either $0$ or $1$ for any $\alpha\in R_+$. Then
\begin{equation}\la{qm}
L_b=\sum_{\pi\in Wb} \,A_\pi\, t(\pi)\,,\qquad A_\pi=\prod_{\genfrac{}{}{0pt}{}{\alpha\in R}{\dpr{\pi, \alpha}>0}}\, 
\mathbf{c}_{\alpha}\,.
\end{equation}

(ii) Let $b\in P^\vee_+$ be quasi-minuscule, i.e. $b=\varphi^\vee$, with $\varphi\in R_+$ the highest root. In this case, $\dpr{\alpha, b}\in\{0, 1\}$ for any $\alpha\in R_+\setminus\{\varphi\}$. Then
\begin{align}\la{qm1}
L_b&=\sum_{\pi\in Wb} \, A_\pi\, (t(\pi)-1)\,,\qquad A_\pi=
\mathbf{c}_{\delta+\pi^\vee}\prod_{\genfrac{}{}{0pt}{}{\alpha\in R}{\dpr{\pi, \alpha}>0}}\, 
\mathbf{c}_{\alpha}\,.
\end{align}
\end{theorem}

\begin{remark}
Strictly speaking, the above expressions for $L_b$ omit an additional constant summand.     
\end{remark}
\begin{remark}
    In the $C^\vee C_n$-case, one takes $L_r=\Res\, f_r$, where $f_r$ is the $r$-th elementary symmetric polynomial of $Y_i+Y_i^{-1}$, $i=1,\dots, n$. 
    Explicit expressions for $L_r$ exist. The simplest one, $L_1$, is known as the {\it Koornwinder operator}. 
\end{remark}

\subsection{DAHAs}\label{daha} In the setting of \ref{bas}, let $P$ be the weight lattice of $R$, defined by $\dpr{P, Q^\vee}=\Z$. The group algebra of the lattice $P$ will be denoted as $\c[X]$. It is spanned by $X^\mu$ with $\mu\in P$, with  $X^\mu X^\nu=X^{\mu+\nu}$. We view $\c[X]$ as a subalgebra of $\c(V)$, by $X^\mu\mapsto e^\mu$.

\begin{defi}
    The algebra $\mathbb H$, generated by $\c[X]$ and $\HH$ viewed as subalgebras of $\D_q(V)*W$, is called the {\it double affine Hecke algebra} (DAHA) of type $R$. 
\end{defi}
Equivalently, $\mathbb H$ is generated by $\c[X]$, $\c[Y]$ and $T_1, \dots, T_n$. Moreover, the elements $X^\mu T_w Y^\lambda$ with $\mu\in P$, $\lambda\in P^\vee$, $w\in W$ form a linear basis in $\mathbb H$. Hence, $\mathbb H$ has the PBW property: a vector space isomorphism $\mathbb H\stackrel{\sim}{\to}  \c[X]\otimes \mathfrak H \otimes \c[Y]$. In the above definition of $\mathbb H$, the commutative subalgebras $\c[X]$ and $\c[Y]$ seem to be of a rather different nature. Nevertheless, one has the following non-trivial result.
\begin{theorem}
   Let $\mathbb H^\vee$ denote the DAHA associated to the dual root system $R^\vee$, i.e. with the r\^oles of $P$ and $P^\vee$ interchanged, and with the same parameters $\tau$. The $\c$-linear mapping $\omega\,:\, \mathbb H\to \mathbb H^\vee$ defined by 
\begin{equation}\la{anti}
 \omega(X^\mu T_w Y^\lambda)=X^{-\lambda}T_{w^{-1}} Y^{-\mu} \quad (\mu\in P,\  \lambda\in P^\vee,\  w\in W)    
\end{equation}
is an anti-isomorphism of algebras.
\end{theorem}
The {\it spherical DAHA} is defined as $e_\tau \mathbb H e_\tau$, where 
\begin{equation}\la{etau}
e_\tau=\frac{1}{\sum_{w\in W}\tau_w^2}\sum_{w\in W} \tau_wT_w
\end{equation}
is the idempotent in the Hecke algebra $\mathfrak H$ corresponding to the one-dimensional character $T_w\mapsto \tau_w$ determined by setting $T_i\mapsto \tau_i$ for $i=1,\dots, n$. It has two commutative subalgebras, $e_\tau \c[X]^W e_\tau$ and $e_\tau \c[Y]^W e_\tau$. 
\begin{prop}
    The map \eqref{resq} restricts to an algebra embedding $\Res\,:\, e_\tau \mathbb H e_\tau \to \D_q(V)^W$.
\end{prop}
This produces a subalgebra $\mathbb B: = \Res(e_\tau \mathbb H e_\tau)$ of $W$-invariant difference operators. It has two commutative subalgebras, $\c[X]^W$ and $\Res(e_\tau \c[Y]^W e_\tau)=\Res(\c[Y]^W)$ constructed in Theorem \ref{qheckthm}. The map \eqref{anti} induces an anti-isomorphism $\omega\,:\,\mathbb B\to \mathbb B^\vee$, exchanging these subalgebras.      

\subsection{Classical case}\la{clcaseq}
The classical limit corresponds to $q=e^c\to 1$, and the procedure is similar to \ref{clcase}. Namely, we set $c=\hbar\beta$, with some fixed $\beta$, and consider the algebra
\begin{equation*}
\Ah*W=\c(V)[t_1^{\pm 1},  \dots , t_n^{\pm 1}][[\hbar]]*W\,,\qquad t_k:=e^{\hbar\beta \partial_{b_k}}\,,
\end{equation*}
where $\{b_k\}$ is a basis for $\Lambda$ (e.g., the fundamental coweights when $\Lambda=P^\vee$). We have 
\begin{equation*}
[t_k, f]=\sum_{l=1}^\infty (\hbar\beta)^l\partial_{b_k}^l(f) t_k\,,\qquad \forall\  f\in\c(V)\,.
\end{equation*}
The algebra $A_\hbar$ is a formal deformation of the algebra $A_0=\c(V)[\Lambda]$ whose elements are finite linear combinations $\sum_{\lambda\in\Lambda}f_\lambda e^{\beta p_\lambda}$, where $f_\lambda\in\c(V)$ and $p_\lambda$ is the classical momentum in direction $\lambda$. The algebra $A_0$ is commutative, with the induced Poisson bracket
$\{e^{\beta p_\lambda}, f\}=\beta\partial_{\lambda}(f)e^{\beta p_k}$. We have an algebra isomorphism (''classical limit'' map)
\begin{equation*}
\eta_0:\ \Ah*W/(\hbar\Ah*W)\to A_0*W\,,\qquad f\mapsto f\,,\ \ t_k\mapsto e^{\beta p_{b_k}}\,,\ \ w\mapsto w\,.
\end{equation*}
The classical Cherednik operators  $Y^{\lambda}_{c}:=\eta_0(Y^\lambda)$ can be defined directly using the classical version of the basic representation, $\beta_c=\eta_0\circ\beta\,:\, \HH\to A_0*W$. For any $f\in\c[Y]^W$, the classical Macdonald--Ruijsenaars Hamiltonian $L_{f, c}=\eta_0(L_f)$ can be obtained as $L_{f, c}=\Res(f(Y_c))$, and the family $\{L_{f, c}\,,\, {f\in\c[Y]^W}\}$ is Poisson-commutative.

\subsection{Nonsymmetric Macdonald polynomials} For brevity, our discussion here is restricted to the settings of \ref{bas}, \ref{daha}. From \eqref{br}, \eqref{br1} it is clear that the algebra $\HH$ preserves the polynomial subspace $\c[X]$ when acting on $\c(V)$. The {\it nonsymmetric Macdonald polynomials} form a basis in $\c[X]$, diagonalising the action of the Cherednik operators. To state the result, we need a particular partial ordering on the weight lattice $P$.

Let $Q_+$ and $P_+$ denote the cones of positive roots and dominant weights, respectively. We write $P_-:=-P_+$. For $\lambda\in P$, denote by $\lambda^{\pm}$ the unique elements of $P_{\pm}\cap W\lambda$, and by $v(\lambda)\in W$ the shortest element such that $v.\lambda=\lambda_-$. For $\lambda, \mu\in P_+$, we say that $\lambda<\mu$ if $0\ne \mu-\lambda\in Q_+$. For $\lambda, \mu\in P$ we say that $\lambda\le \mu$ if either $\lambda^+<\mu^+$ or $\lambda^+=\mu^+$ and $v(\lambda)\ge v(\mu)$ with respect to the Bruhat order on $W$. 
Note that any antidominant $\lambda\in P_-$ is highest in its $W$-orbit. 

\begin{prop}
    For generic parameters $q=e^c$ and $\tau$, and for any $\mu\in P$, there is a unique  $E_\mu\in\c[X]$ of the form $E_\mu=X^\mu+\mathrm{l.o.t.}$ which is a common eigenfunction of the Cherednik operators: 
    \begin{equation}
        Y^\lambda(E_\mu)=\gamma_{\lambda,\mu}E_\mu\quad  (\mu\in P, \ \lambda\in P^\vee)
    \end{equation}
    with $\gamma_{\lambda, \mu}\in\c$. 
\end{prop}
The polynomials $E_\mu$ ($\mu\in P$) are called the monic {\it nonsymmetric Macdonald polynomials}. Their coefficients are rational functions of $q, \tau$. 
The {\it symmetric Macdonald polynomials} are obtained from $E_\mu$ by applying the symmetriser \eqref{etau}. They form a basis of $\c[X]^W$ that diagonalises the action of the operators $L_f$ from Theorem \ref{qheckthm}.

\subsection{Affine $q$-KZ equations} The DAHAs can be used to construct a $q$-analogue of the KZ equations \eqref{kz}. Pick a module $\tau$ over the affine Hecke algebra, $\HH$, and consider the space $M(\tau):=\c(V)\otimes\tau $. It will be viewed as $\tau$-valued meromorphic functions on $V$, acted upon by $(\D_q(V)*W)\otimes \HH$. For $a\in\D_q(V)$, $w\in W$, $h\in\HH$, we abbreviate the action of $aw\otimes h$ on $M(\tau)$ as $awh$. (Note that the actions of $\D_q(V)*W$ and $\HH$ on $M(\tau)$ commute.) 
Recall that the group algebra of $\Wh$ sits inside $\D_q(V)*W$ by \eqref{hatin}, and together with $f\in \c(V)$ it generates $\D_q(V)*W$.
\begin{prop}\la{nabla}
The assignment 
\begin{align*}
\nabla(f)&=f\,,\\
\nabla(s_i)&=(\mathbf{c}_i)^{-1}s_iT_i+\frac{\mathbf{c}_i-\tau_i}{\mathbf{c}_i}s_i\,,\\
\nabla(\omega)&=\omega T_\omega
\end{align*}
for $f\in\c(V)$, $i=0, \dots, n$, $\omega\in\Omega$ extends uniquely to an algebra map $\nabla\,:\, \D_q(V)*W\to \End_\c M(\tau)$, that is, an action of $\D_q(V)*W$ on $M(\tau)$. 
\end{prop}
To arrive at the above definition of $\nabla$, one argues similarly to the calculation of the KZ connection \eqref{conn}. Namely, one first uses the embedding $\HH\to \mathbb H$ and the PBW property of $\mathbb H$ to induce a structure of an $\mathbb H$-module on $\c[X]\otimes \tau$. A localisation $\mathbb H_{\mathrm{reg}}$ to a suitable subset $\vreg\subset V$ identifies $\mathbb H_{\mathrm{reg}}\cong \D_q[\vreg]*W$, which in its turn induces a structure of a $\D_q[\vreg]*W$-module on $M=\c[\vreg]\otimes \tau$. A direct calculation 
then shows that $s_i$ acts on $M$ by the above formula (for $f\in\c[\vreg]$ and $\omega\in\Omega$ this is obvious). By analytic continuation, the same action is well defined on $M(\tau)=\c(V)\otimes \tau$.

\begin{defi}
The affine $q$-KZ equations (with values in a $\HH$-module $\tau$) is the following system of equations for $F\in M(\tau)$: 
\begin{equation}\la{aqkz}
 \nabla(t(\lambda))F=F\qquad \quad\forall\ \lambda\in P^\vee\,.
\end{equation}
The action of $\nabla(w)$, $w\in W$ makes the solution space to \eqref{aqkz} into a $W$-module. 
 \end{defi}
A convenient way of expressing $\nabla(t(\lambda))$ and, more generally, $\nabla(w)$ for any $w\in\Wh$ is by
\begin{equation*}
    \nabla(w)=C_w w\,,\quad C_w\in\c(V)\otimes\HH\,.
\end{equation*}
Here $\{C_w,\ w\in\Wh\}$ satisfy the cocycle condition
\begin{equation}\la{coc}
    C_{ww'}=C_w\,(w.C_{w'})\qquad \forall\ w,w'\in \Wh\,.
\end{equation}
(The action of $w$ on $C_{w'}\in\c(V)\otimes \HH$ is by $w\otimes 1$.) From Proposition \ref{nabla} we find
\begin{equation*}
    C_{s_i}=(\mathbf{c}_i)^{-1}T_i+\frac{\mathbf{c}_i-\tau_i}{\mathbf{c}_i}=\frac{T_i^{-1}-e^{a_i}T_i}{\tau_i^{-1}-\tau_ie^{a_i}}\,,\qquad C_{\omega}=T_\omega\,,
\end{equation*}
so together with \eqref{coc} this uniquely determines $C_w$. 

\subsection{Trigonometric Dunkl operators, Calogero--Moser--Sutherland system, and degenerate DAHA} 
The rational Cherednik algebra can be regarded as a rational version of DAHA. There is also an intermediate version, referred to as the {\it degenerate} DAHA. In that version, the affine Hecke algebra $\HH$ is replaced with the {\it graded Hecke algebra}, breaking the symmetry between $\c[X]$, $\c[Y]$.  

We begin in the setting of \ref{3.1}. For a $W$-invariant multiplicity function $k\,:\, R\to \c$, 
set 
\begin{equation}\label{rho}
\rho_k=\frac12\sum_{\alpha\in R_+} k_\alpha \alpha\,.    
\end{equation}
\begin{defi}[Dunkl--Cherednik operator]
For $\xi\in V$, define
\begin{equation}\label{dch}
    T_\xi=\partial_\xi+\sum_{\alpha\in R_+}k_\alpha\dpr{\alpha, \xi} \frac{1}{1-e^{-\alpha}}(1-s_\alpha)-\dpr{\rho_k, \xi}\,.
\end{equation}
\end{defi}
This is a trigonometric analogue of the operators \eqref{dut}. Their key property is, again, the commutativity: $[T_\xi, T_\eta]=0$ for $\xi, \eta\in V$, hence the mapping $V\ni\xi\mapsto T_\xi$ extends to an algebra map
\begin{equation}\label{tp}
    \c[V^*]\to \D(V)*W,\qquad q\mapsto T_q\,.
\end{equation}

\begin{prop}\label{trigcher}
The map $q\mapsto L_q:=\Res\,T_q$, $q\in \c[V^*]^W$ defines an algebra embedding $\c[V^*]^W\to\D(V)^W$.     
\end{prop}
By a direct calculation, 
\begin{equation}\label{cmsrad}
    L_{\dpr{\xi, \xi}}= \Delta + \sum_{\alpha\in R_+} k_\alpha\frac{1+e^{-\alpha}}{1-e^{-\alpha}}\partial_\alpha + \dpr{\rho_k, \rho_k}\,. 
\end{equation}
This is a trigonometric analogue of \eqref{cmrad}. Using a suitable gauge transformation, it can be transformed into the hyperbolic Calogero--Moser--Sutherland Hamiltonian 
\begin{equation}\label{cms}
    L= \Delta - \sum_{\alpha\in R_+} k_\alpha(k_\alpha-1)\frac{\dpr{\alpha, \alpha}}{4\sinh^2\frac{\alpha}{2}}\,. 
\end{equation}

The Dunkl--Cherednik operators are manifestly not $W$-equivariant. An alternative definition is 
\begin{equation}\label{dh}
   S_\xi=\partial_\xi+\frac12 \sum_{\alpha\in R_+}k_\alpha\dpr{\alpha, \xi} \frac{1+e^{-\alpha}}{1-e^{-\alpha}}(1-s_\alpha)\,. 
\end{equation}
These are called the {\it Dunkl--Heckman} operators. They are $W$-equivariant, but do not commute. Nevertheless, they can be used to show the integrability of the Hamiltonian \eqref{cmsrad}.
\begin{prop}\label{trigheck}
Given $v\in V$ and $r\in\N$, define 
\begin{equation*}
 L_{v, r}=\Res\,\left(\sum_{\xi\in Wv} (S_\xi)^r\right)\,.   
\end{equation*}
 Then the operators $L_{v, r}$ for $v\in V$, $r\in\N$ form a commutative family, containing the Hamiltonian \eqref{cmsrad}.
\end{prop}

\medskip

\begin{example}
    For $W=S_n$, the positive roots are $\alpha=e_i-e_j$ with $i<j$, and $k_\alpha=k$ for all $\alpha$. The Dunkl--Heckman operators $S_i=S_{e_i}$ take the following form:
 \begin{equation*}
   S_i=\frac{\partial}{\partial x_i}+\frac{k}{2} \sum_{j \ne i}^n \frac{e^{x_i}+e^{x_j}}{e^{x_i}-e^{x_j}}(1-s_{ij})\,. 
\end{equation*}   
Written in coordinates $z_i=e^{x_i}$, these become
 \begin{equation*}
   S_i=z_i\frac{\partial}{\partial z_i}+\frac{k}{2} \sum_{j \ne i}^n \frac{{z_i}+{z_j}}{{z_i}-{z_j}}(1-s_{ij})\,. 
\end{equation*}   
The operators 
\begin{equation}\label{lr}
L_r=\Res\,\left(\sum_{i=1}^n S_i^r\right), \qquad r=1, \dots, n     
\end{equation}
generate the algebra of commuting quantum Hamiltonians in this case. A slightly different family of operators was introduced by Polychronakos:
\begin{equation}\label{po}
   \pi_i=z_i\frac{\partial}{\partial z_i}+{k}\sum_{j \ne i}^n \frac{{z_i}}{{z_i}-{z_j}}(1-s_{ij})\,. 
\end{equation} 
They satisfy the commutation relations
  $[\pi_i, \pi_j]=-k(\pi_i-\pi_j)s_{ij}$  
which can be used to show that the operators
\begin{equation}\label{ir}
    I_r=\Res\,\left(\sum_{i=1}^n \pi_i^r\right)\,,\qquad r=1, \dots, n    
\end{equation}
pairwise commute. Although $I_r\ne L_r$ in general, the algebras generated by $L_r$ and $I_r$ coincide.
\end{example}


\medskip

Let $\Wh^\vee=\Wh(R^\vee)=W\rtimes P$ be the extended affine Weyl group for the dual root system, and $\Omega^\vee\cong P/Q$ be the abelian subgroup of elements of length zero. The group algebra of $\Wh^\vee$ can be realised inside $\D(V)*W$ by sending $W\ni w\mapsto w$ and $P\ni \lambda\mapsto e^\lambda$ (viewed as a function $x\mapsto e^{\dpr{\lambda, x}}$).

\begin{defi}
    The subalgebra $\mathbf H_k$ of $\D(V)*W$, generated by $w$, $e^\lambda$, $T_\xi$ ($w\in W$, $\lambda\in P$, $\xi\in V$), is called the {\it degenerate DAHA} associated to the root system $R$ and multiplicity function $k$.
\end{defi}
The algebra $\mathbf H_k$ has the PBW property: a vector space isomorphism $\mathbf H_k\stackrel{\sim}{\to}  \c[V^*]\otimes \c\Wh^\vee$. The elements $T_pe^\lambda w$ ($p\in \c[V^*]$, $\lambda\in P$, $w\in W$) from a linear basis in $\mathbf H_k$. 

\medskip

Another realisation of the same algebra is constructed inside $\D_q(V)*W$ (with the lattice $\Lambda=P$). Recall that one can realise $\Wh^\vee$ inside $\D_q(V)*W$, sending $W\ni w\mapsto w$, $P\ni \lambda\mapsto t(\lambda)=e^{c\partial_\lambda}$. This representation will be denoted $\pi_0$ and it can be deformed as follows.
Introduce the elements
\begin{equation}
    r_i=s_i-\frac{k_i}{a_i^\vee}(1-s_i)\,,\quad (i=0,\dots, n)\,,\quad s_i=s_{a_i^\vee}\,.
\end{equation}
Here $a_0^\vee=\delta-\psi^\vee$, $a_1^\vee, \dots, a_n^\vee$ are simple roots of $\Ra^\vee$, $k_i=k_{a_i}$ ($i=1, \dots, n$), $k_0=k_{\psi}$, with $\psi$ the highest short root of $R$. This determines a representation of $\Wh^\vee$ by 
\begin{equation}
  \pi_k\,:\, \c\Wh^\vee\to \D_q(V)*W\,,\quad s_i\mapsto r_i\,,\quad  \Omega^\vee\ni \omega\mapsto \omega\quad (i=0, \dots, n)\,.
\end{equation}
(Note that $\pi_k(\omega)=\pi_0(\omega)$ for $\omega\in\Omega$.) As a consequence, we have a realisation of the lattice $P$ by $\lambda\mapsto \pi_k(\lambda)$. The difference operators $\pi_k(\lambda)$ represent a rational limit of the Cherednik operators $Y^\lambda$. 

Let us identify $\c[V^*]\stackrel{\sim}{\to} \c[V]$, $q\mapsto q^*$ using the Euclidean form $\dpr{\cdot, \cdot}$.
\begin{prop}\label{dualrep}
    The algebra $\mathbf H_k$ is isomorphic to the subalgebra $\mathbf H'_k$ of $\D_q(V)*W$ generated by $\c[V]$, $r_i$ ($i=0, \dots, n$), and $\omega\in\Omega^\vee$. The isomorphism $\mathbf H_k \to \mathbf H_k'$ sends $W\ni s_i\mapsto r_i$ ($i=1,\dots,n$), $e^\lambda\mapsto \pi_k(\lambda)$, $T_q\mapsto q^*$. 
\end{prop}

\subsection{Historical comments}
Most of the results here are due to Cherednik. For the construction of Cherednik operators and Theorems \ref{bre}, \ref{qheckthm} as well as for the introduction of DAHAs, see \cite{C1, C2}. The duality (Theorem \ref{dua}) is stated in \cite{C3} and proved in \cite{M03}. For the $C^\vee C_n$-case, the basic representation is due to Noumi \cite{No}, see also Sahi's work on the corresponding DAHA theory and duality \cite{Sa}, as well as \cite{St1}. Note that in the $\GL_n$-case, the operators \eqref{yi} appeared in \cite{BGHP} in connection with long-range spin chains and the Yangian. Another $q$-analogue of Dunkl operators for the $\GL_n$-case, different from \eqref{yi}, was proposed in \cite{BF}. It can be viewed as a special case of a more general family introduced recently in the theory of cyclotomic DAHAs \cite{BEF}.

The operators \eqref{ru} are equivalent to the Hamiltonians found by Ruijsenaars \cite{R87}. The Macdonald operators \eqref{qm}, \eqref{qm1} first appeared as part of Macdonald's theory of (symmetric) Macdonald polynomials in \cite{M87}. For the $C^\vee C_n$-case, a second order difference operator was found by Koornwinder \cite{Ko}, who also introduced the corresponding generalisation  of the Macdonald polynomials (Koornwinder polynomials). Explicit expressions for higher order difference operators in that case were found by van Diejen \cite{vD}. 

Nonsymmetric Macdonald polynomials were introduced in \cite{M96, C3}. Our setting is not the most general: it corresponds to the case of {\it non-twisted} affine root systems. For comprehensive accounts of the theory of DAHAs and Macdonald--Koornwinder polynomials, see \cite{M03, C3a, St2}. The affine $q$-KZ equations appeared in \cite{C2, C94a}, our account follows closely \cite{St3} where one also finds a discussion of possible choices for the $\HH$-module $\tau$. As explained in \cite[1.3.2]{C3a}, in the $\GL_n$-case one can reproduce the $q$-KZ equations of Smirnov and Frenkel--Reshetikhin \cite{Sm, FR}. For the links between nonsymmetric Macdonald polynomials and solutions to the qKZ system, see \cite{KT,St3}. In the $\GL_n$-case, Macdonald polynomials and Macdonald operators can be defined at the level of symmetric functions (of infinite number of variables). Such approach can also be applied to trigonometric Dunkl and Cherednik operators, see \cite{SV1,NS17}.

The Dunkl--Cherednik operators \eqref{dch} and Proposition \ref{trigcher} are due to Cherednik \cite{C1, C94}. The $W$-equivariant version \eqref{dh} and Proposition \ref{trigheck} are due to Heckman \cite{He2}. The operators \eqref{po} appeared in \cite{P}; for the precise relationship between the operators \eqref{lr} and \eqref{ir}, see \cite[Prop. 5.2]{SV2} .

The degenerate DAHAs were introduced in \cite{C97, C98}. We followed Opdam's lectures \cite{O2}, where a nice account of the theory and its applications to harmonic analysis and special functions on root systems can be found. The ``dual'' basic representation from Proposition \ref{dualrep} is discussed in \cite[1.6.4]{C3}.

\section{Elliptic Dunkl and Cherednik operators and Calogero--Moser and Ruijsenaars systems}\la{ecase}

\subsection{Elliptic Dunkl operators}\label{edo} In the setting of Section \ref{2.1}, let $W$ be a Weyl group 
with a root system $R$ and a multiplicity function $k$. Fix $\tau\in\c$ with $\mathrm{Im}\,\tau>0$. The notation $\sigma_\mu(z)$ will be used throughout for
\begin{equation}\la{sigm}
\sigma_\mu(z) =\frac{\theta(z-\mu)\theta'(0)}{\theta(z)\theta(-\mu)}\,,\qquad \mu, z\in \c\,. 
\end{equation}
Here $\theta(z)=\theta_1(z |\tau)$ is the odd Jacobi theta function associated with the elliptic curve $\c/\Z+\Z\tau$. 

For $\lambda\in V$, the {\it elliptic Dunkl operators} are the following elements of $\D(V)*W$:
\begin{equation}
\label{edu} y_\xi =
\hb\partial_\xi-\sum_{\alpha\in R_+}
k_\alpha \langle\alpha, \xi\rangle \sigma_{\langle\alpha^\vee, \lambda\rangle}(\langle\alpha, x\rangle)s_\alpha\ , \quad \xi \in V\ .
\end{equation}
The auxiliary {\it spectral variable} $\lambda$ is a distinctive feature of the elliptic case, 
%
and we write $y_\xi(\lambda)$ when need to emphasize the dependence on $\lambda$. Note that as a function of $\lambda$, \eqref{edu} has poles along the hyperplanes $\dpr{\alpha^\vee, \lambda}=m+n\tau$ with $m,n\in\Z$. Once again, two main properties of the Dunkl operators are their commutativity and equivariance: 
for all $\,\xi, \eta \in V\,$ and $ w \in W $,
\begin{equation}\la{eduprop}
 y_{\xi}\,y_{\eta} = y_{\eta}\,y_{\xi}\,,\qquad \,w\,y_\xi(\lambda) =
y_{w\xi}(w\lambda)\,w\,.
\end{equation}
Note that in the second relation the group action now changes both $\xi$ and $\lambda$. As before, the assignment $\,\xi \mapsto y_\xi\,$
extends to an algebra map
\begin{equation}\label{ehom}
\c[V^*] \to \D(V)*W \ ,\quad q\mapsto q(y)\,.
\end{equation}
%

By taking suitable symmetric combinations of the elliptic Dunkl operators, one can construct commuting Hamiltonians of the elliptic Calogero--Moser system.
We first illustrate the procedure for the quadratic Hamiltonian. Similarly to Example \ref{cmd}, we calculate $\dpr{y, y}=y_{1}^2+\dots +y_{n}^2$ to find 
\begin{equation}
\dpr{y, y}=\hb^2\Delta-\hb\sum_{\alpha\in R_+}k_\alpha\langle\alpha, \alpha\rangle\sigma'_{\langle\alpha^\vee, \lambda\rangle}(\langle\alpha, x\rangle)s_\alpha+
\sum_{\alpha\in R_+} k^2_\alpha\langle\alpha, \alpha\rangle \left(\wp(\dpr{\alpha^\vee, \lambda})-\wp(\dpr{\alpha, x}) \right)\,.
\end{equation} 
Here $\sigma'_\mu(z)=\frac{d}{dz}\sigma_\mu(z)$, and $\wp(z)$ is the Weierstrass $\wp$-function with periods $1, \tau$. 

The resulting expression is singular at $\lambda=0$, but can be regularised by subtracting a $\lambda$-dependent term. Using that 
$\lim_{\mu\to 0}\sigma'_\mu(z)=-\wp(z)-2\zeta(\frac{1}{2})$, one finds that
\begin{equation}
 \dpr{y, y}-\sum_{\alpha\in R_+} \frac{k^2_\alpha \langle\alpha, \alpha\rangle}{\langle\alpha^\vee, \lambda\rangle^2} \ \to\  \hb^2\Delta-\sum_{\alpha\in R_+} k_\alpha(k_\alpha-\hb s_\alpha)\langle\alpha, \alpha\rangle \wp(\langle\alpha, x\rangle)+C\quad\text{as $\lambda\to 0$}\,, 
\end{equation}
with $C=2\zeta(\frac12) \hb\sum_{\alpha\in R_+} k_\alpha\dpr{\alpha, \alpha} s_\alpha$. Applying the map \eqref{res}, we define
\begin{equation}\label{ecal}
    L_{\dpr{\xi,\xi}}:=\Res \lim_{\lambda\to 0}\, \left(  \dpr{y, y}-\sum_{\alpha\in R_+} \frac{k^2_\alpha \langle\alpha, \alpha\rangle}{\langle\alpha^\vee, \lambda\rangle^2} \right)\,.
\end{equation}
Up to a constant, this produces the Hamiltonian of the quantum elliptic Calogero--Moser system,
\begin{equation}\label{elcm}
    L=\hb^2\Delta-\sum_{\alpha\in R_+} k_\alpha(k_\alpha-\hb)\langle\alpha, \alpha\rangle \wp(\langle\alpha, x\rangle)\,.
\end{equation}

\subsection{Integrability of the elliptic Calogero--Moser system}\label{iecm} 
Commuting Hamiltonians for the operator \eqref{elcm} are constructed as follows. First, extend the algebra map \eqref{ehom} by allowing polynomials with $\lambda$-dependent coefficients:
\begin{equation}\label{ehomex}
\c(V)\otimes \c[V^*] \to \D(V)*W \ ,\quad f\mapsto f(\lambda, y)\,.
\end{equation}
Next, recall the subalgebra $\mathcal A^c\subset A_0^W$ of classical Hamiltonians $L_q^c$ constructed in Theorem \ref{heckthmc}. Define $(\mathcal A^{c})^\vee$ to be such an algebra constructed for the root system $R^\vee$ and multiplicities $k_{\alpha^\vee}:=k_\alpha\dpr{\alpha, \alpha}/2$. Write $f_q$, $q\in\c[V^*]$ for the elements of $(\mathcal A^c)^\vee$.  
\begin{theorem}\label{efmv}
For $q\in\c[V^*]^W$, consider $f_q\in (\mathcal A^{c})^\vee\subset A_0^W$. Identify $A_0$ with $\c(V)\otimes 
\c[V^*]$ in\eqref{ehomex} and obtain $f_q(\lambda, y)\in \D(V)*W$ by applying \eqref{ehomex} to $f_q$. 

$(1)$ The elements $f_q(\lambda, y)$ are regular near $\lambda=0$ and so have a well-defined limit as $\lambda\to 0$. 

$(2)$ Setting $L_q:=\Res \lim_{\lambda\to 0}\,f_q(\lambda, y)$ defines an algebra embedding $\c[V^*]^W\to \D(V)^W$, $q\mapsto L_q$.
\end{theorem}
Our calculation in \eqref{ecal} is a particular example of this construction for $q=\dpr{\xi, \xi}$ and $f_q$ being the classical rational Calogero--Moser Hamiltonian \eqref{cmcl}.

As a corollary, this demonstrates that the quantum system \eqref{elcm} is completely integrable as it admits $n=\dim V$ commuting Hamiltonians $L_q$. 
The classical limit of $L_q$ is obtained by replacing the operators \eqref{edu} by their classical limit: 
\begin{equation}
\label{educ} y_\xi^c =
p_\xi-\sum_{\alpha\in R_+}
k_\alpha \langle\alpha, \xi\rangle \sigma_{\langle\alpha^\vee, \lambda\rangle}(\langle\alpha, x\rangle)s_\alpha\ , \quad \xi \in V\ .
\end{equation}
Hence, the above theorem also produces a family of Poisson-commuting Hamiltonians for the classical system described by
\begin{equation}\label{elcmcl}
    L^c=\dpr{p, p}-\sum_{\alpha\in R_+} k_\alpha^2\langle\alpha, \alpha\rangle \wp(\langle\alpha, x\rangle)\,.
\end{equation}

\subsection{$BC_n$-case}\la{inoz} 
A similar method applies to the $BC_n$-version of the Hamiltonian \eqref{elcm}, describing the Inozemtsev system:
\begin{equation}
L=\hb^2\Delta-2k(k-\hb)\sum_{i<j}^n (\wp(x_{i}-x_j)+\wp(x_i+x_j))-\sum_{i=1}^n\sum_{r=0}^3 g_r(g_r-\hb)\wp(x_i+\omega_r)\,.    
\end{equation}
Here $\omega_r$ are the elliptic half-periods and $k, g_0, g_1, g_2, g_3$ are the multiplicities. 
The group $W$ is of type $B_n$, with reflections $s_i$ (changing sign of $x_i$), $s_{ij}$ (permuting $x_i, x_j$), and $s_{ij}^+=s_is_{ij}s_i$. The Dunkl operators are  
\begin{equation*}
y_i=\hb\partial_i-v_{\lambda_i}(x_i)s_i-k\sum_{j\ne i}\left(\sigma_{\lambda_i-\lambda_j}(x_i-x_j)s_{ij}+\sigma_{\lambda_i+\lambda_j}(x_i+x_j)s_{ij}^+\right)\,,\qquad i=1\dots n\,.
\end{equation*}
Here $\lambda=(\lambda_1, \dots, \lambda_n)$ are the spectral variables, and 
\begin{equation}\la{vmu}
v_\mu (z)=v_\mu(z; g_0, g_1, g_2, g_3)=\sum_{r=0}^3 g_r\sigma^r_{2\mu}(z)\,,\qquad \sigma^r_\mu(z):=\frac{\theta_{r+1}(z-\mu)\theta_1'(0)}{\theta_{r+1}(z)\theta_1(-\mu)}\,,
\end{equation}
where $\theta_r(z)=\theta_r(z|\tau)$, $r=0\dots 3$ are the Jacobi theta functions, with $\theta_4(z):=\theta_0(z)$.

\subsection{Elliptic difference case}\la{eqcase} Let us now introduce elliptic Cherednik operators and use them to show the integrability of the elliptic Ruijsenaars system and its generalisations for other root systems.  
We will refer to these systems as \emph{generalised Ruijsenaars systems}. The main tool is elliptic functional $R$-matrices.

The setting is the same as in \ref{3.1}: $R\subset V$ is a reduced, irreducible root system with Weyl group $W$, $k$ is a $W$-invariant multiplicity function, $\Ra$ is the associated affine root system with a chosen basis $a_0, \dots, a_n$, $\Wh=W\rtimes t(P^\vee)$ is the extended affine Weyl group with translations $t(\lambda)=e^{c\partial_\lambda}$. 

For $\aalpha=\alpha+m\delta\in\Ra$, define $R$-matrices $\RR(\aalpha)$ to be the following elements of $\D_q(V)*W$ :
\begin{equation}\la{raa}
\RR(\aalpha)=\sigma_{{k}_\alpha}(\aalpha)-\sigma_{\dpr{\alpha^\vee,\,\xi}}(\aalpha)s_{\aalpha}\,,
\end{equation}
where we now denote the spectral variables as $\xi\in V$ rather than $\lambda$. The notation $\sigma_{\mu}(z)$ is as in \eqref{sigm}. 
These $R$-matrices are unitary in the sense that
\begin{equation}\la{uni}
\RR(\alpha) \RR(-\alpha)=\wp({k}_\alpha)-\wp(\dpr{\alpha^\vee, \xi})\,.
\end{equation} 

\begin{defi}\la{rwdef}
Define a set $\{\RR_w\,| w\in\Wh\}$ by taking a reduced decomposition $w=s_{i_1}\dots s_{i_l}\omega$, $\omega\in\Omega$ and setting
\begin{equation}\la{rw}
\RR_w=\RR(\alpha^{1})\dots \RR(\alpha^{l})\,,\quad\text{where}\quad \alpha^{1}=a_{i_1},\ \alpha^{2}=s_{i_1}(a_{i_2})\,,\ \dots,\ \alpha^{l}=s_{i_1}\dots s_{i_{l-1}}(a_{i_l})\,. 
\end{equation}
In particular, we have $\RR_{s_i}=\RR(a_i)$, $i=0 \dots n$, and $\RR_\omega=1$ for $\omega\in\Omega$. {\it Elliptic Cherednik operators} are defined as $Y^b=\RR_{t(b)}\,t(b)$, $b\in P^\vee_+$. 
\end{defi}

\begin{theorem}
\la{yb}
$(1)$ The elements $\RR_w$ do not depend on the choice of a reduced decomposition for $w$;
$(2)$ $Y^b$, $b\in P^\vee_+$ are pairwise commuting elements of $\D_q(V)*W$. 
\end{theorem}
The proof is based on the fact that $\RR(\aalpha)$ satisfy the {\it affine Yang--Baxter equations} associated to the root system $\Ra$.

The commuting quantum Hamiltonians are obtained from the operators $Y^b$ in the following way. Recall the map \eqref{resq} and the vector $\rho_k$ \eqref{rho}.
\begin{theorem}
\la{emr} Given $b\in P_+^\vee$, set $\xi=-\rho_k$ and let
$L_b=\Res\, Y^b$. Then $L_b$ is $W$-invariant, 
and the difference operators $L_b$, $b\in P^\vee_+$ form a commutative family in $\D_q(V)^W$.
\end{theorem}
This result seems surprising as one does not need to take symmetric combinations of $Y^\lambda$ to produce $L_b$. 
The commuting difference operators $L_b$ define an integrable system. For the $\GL_n$-case this coincides with the Ruijsenaars system (see \eqref{emr1} below), for other cases this gives its generalisation. 

The operators $L_b$ are complicated in general, but some admit an explicit description similar to the Macdonald operators in Theorem \ref{mo}. 
\begin{theorem}
\la{lb}  
(i) Let $b$ be a minuscule coweight, so that $\dpr{\alpha, b}$ is either $0$ or $1$ for any $\alpha\in R_+$. Then we have
\begin{equation}
L_b=\sum_{\pi\in Wb} \,A_\pi t(\pi)\,,\qquad A_\pi=\prod_{\genfrac{}{}{0pt}{}{\alpha\in R}{\dpr{\pi, \alpha}>0}}\, \sigma_{{k}_\alpha}(\alpha)\,.
\end{equation}

(ii) Let $b$ be a quasi-minuscule coweight of the form $b=\varphi^\vee$, with $\varphi\in R_+$ the highest root. In this case, $\dpr{\alpha, b}\in\{0, 1\}$ for any $\alpha\in R_+\setminus\{b\}$. Then
\begin{align}\la{lbq1}
L_{b}&=\sum_{\pi\in Wb} \, (A_\pi t(\pi)-B_\pi)\,,\qquad A_\pi=\sigma_{{k}_{\varphi}}(\pi^\vee+\delta)
\prod_{\genfrac{}{}{0pt}{}{\alpha\in R}{\dpr{\pi, \alpha}>0}}\, \sigma_{{k}_\alpha}(\alpha)\,,
\\\la{lbq2}
B_\pi&=\sigma_{\dpr{\varphi^\vee, -\rho_k}}(\pi^\vee+\delta)
\prod_{\genfrac{}{}{0pt}{}{\alpha\in R}{\dpr{\pi, \alpha}>0}}\, \sigma_{{k}_\alpha}(\alpha)\,.
\end{align}
In these formulas the roots are viewed as affine-linear functions, so, for example,  $\sigma_{{k}_\alpha}(\alpha+\delta)=\sigma_{{k}_\alpha}(\dpr{\alpha, x}+c)$.
\end{theorem}




\subsection{$\GL_n$-case}\la{arelel}
The setting is similar to \ref{aq}: we take $V=\c^n$, with the orthonormal coordinates $x_1, \dots, x_n$ and the standard action of $W={S}_n$. We set $\Lambda=\sum_{i=1}^n \Z\epsilon_i$, and consider the algebra of difference operators ${\D}_q(V)$ for the lattice $\Lambda$. 
We have one constant ${k}_\alpha={k}$ for all $\alpha\in R$. For $\alpha=\epsilon_i-\epsilon_j$, $i\ne j$, the $R$-matrices \eqref{rdef} take the form 
\begin{equation}\la{rije}
\RR_{ij}=\sigma_{{k}}(x_i-x_j)-\sigma_{\xi_i-\xi_j}(x_i-x_j)s_{ij}\,.
\end{equation}
They satisfy the Yang--Baxter relations,
$\RR_{ij}\RR_{ik}\RR_{jk}=\RR_{jk}\RR_{ik}\RR_{ij}$ (for $i\ne j\ne k$). 
The elliptic Cherednik operator $Y_1:=Y^{\epsilon_1}$ can be calculated from Definition \ref{rwdef} to give 
\begin{equation}\la{yie}
Y_1=\RR_{12}\RR_{13}\dots \RR_{1n}\,t(\epsilon_1)\,.
\end{equation} 
The elliptic Ruijsenaars operator $L_1=L_{\epsilon_1}$ is 
\begin{equation}\la{emr1}
L_1=\sum_{i=1}^n \prod_{j\ne i}^n \sigma_{k}(x_i-x_j)\,t(\epsilon_i)\,.
\end{equation}  
Up to a gauge transformation, this is the quantum Hamiltonian of the elliptic Ruijsenaars system. 
Other fundamental weights $b=\epsilon_1+\dots+\epsilon_r$ produce
\begin{equation}
\label{eru} L_r=\sum_{\genfrac{}{}{0pt}{}{I\subset
\{1,\dots,n\}}{|I|=r}} \prod_{\genfrac{}{}{0pt}{}{i\in I} {j\notin
I}} 
\sigma_k(x_i-x_j)\,\prod_{i\in I} t(\epsilon_i)\,.
\end{equation}

\subsection{$C^\vee C_n$-case}\la{ecc}
We are in the setting of Section \ref{cc}, in particular, the lattice is $\Lambda=\sum_{i=1}^n \Z\epsilon_i$. The $R$-matrices in this case depend on the spectral variables $\xi\in V$ and coupling constants $\mu$, $\nu$, 
$\ov{\nu}$, $g=(g_i)$, $\ov{g}=(\ov{g}_i)$ ($i=0\dots 3$), and are as follows:
\begin{align}\label{cel1}
\RR(\aalpha) &=\sigma_\mu(\aalpha)-\sigma_{\dpr{\alpha^\vee, \xi}}(\aalpha)s_{\aalpha}\,\quad &&\text{for\ }\aalpha=k\delta\pm\epsilon_i\pm\epsilon_j\quad(k\in\Z,\ i\ne j)\,, \\\la{cel2}
\RR(\aalpha) &= v_{\nu, g}(\aalpha/2) -v_{\dpr{\alpha^\vee, \xi}, g}(\aalpha/2)s_{\aalpha}\,\quad &&\text{for\ }\aalpha=2k\delta\pm 2\epsilon_i\quad(k\in\Z)\,,\\\la{cel3}
\RR(\aalpha) &= {v}_{\ov{\nu}, \ov{g}}(\aalpha/2) -{v}_{\dpr{\alpha^\vee, \xi}, \ov{g}}(\aalpha/2)s_{\aalpha}\,\quad &&\text{for\ }\aalpha=(2k+1)\delta\pm 2\epsilon_i\quad(k\in\Z)\,.
\end{align}
In these formulas, $v_{\nu, g}(z)=v_\nu(z; g_0, g_1, g_2, g_3)$ is the function \eqref{vmu}. 

We can now define the elements $\RR_w$ and $Y^b=R_{t(b)}t(b)$ in the same way as in \ref{rwdef} (note that the group $\Omega$ is trivial in this case), and Theorem \ref{yb} remains valid in this setting. 
For example, we have the following expression for $Y_1:=Y^{\epsilon_1}$:
\begin{align*}
Y_1=\RR (\epsilon_1-\epsilon_{2})\RR (\epsilon_1-\epsilon_{3})\dots \RR (\epsilon_1-\epsilon_n)\RR (2\epsilon_1)\RR (\epsilon_1+\epsilon_n)\dots \RR (\epsilon_1+\epsilon_{2})\RR (\delta+2\epsilon_1)t(\epsilon_1)\,.
\end{align*}

\begin{theorem}
\la{cemr} Specialise $\xi$ to $\xi=(\xi_1,\dots, \xi_n)$ with $\xi_i=-\nu-(n-i)\mu$.

$(1)$ Given $b\in \Lambda_+$, let $L_b=\Res_q Y^b$. Then the operators $L^b$ are commuting, $W$-invariant difference operators.

$(2)$ Let $b=\epsilon_1$. Then
\begin{align}
\la{clb1}
L_{b}&=\sum_{\pi\in Wb} \, (A_\pi t(\pi)-B_\pi)\,,\qquad A_\pi=v_{\nu, g}(\pi){v}_{\ov{\nu}, \ov{g}}(\pi+\delta/2)\,
\prod_{\genfrac{}{}{0pt}{}{\alpha\in R}{\dpr{\pi, \alpha}=1}}\, \sigma_{\mu}(\alpha)\,,
\\
\la{clb2}
B_\pi&=v_{\nu, g}(\pi)
{v}_{-\nu-(n-1)\mu, \ov{g}}
(\pi+\delta/2)
\prod_{\genfrac{}{}{0pt}{}{\alpha\in R}{\dpr{\pi, \alpha}=1}}\, \sigma_{\mu}(\alpha)\,.
\end{align}
\end{theorem}
The operator \eqref{clb1}--\eqref{clb2} is called the Van Diejen's Hamiltonian. It contains $11$ parameters $\mu, \nu, \ov{\nu}, g_i, \ov{g}_i$, but multiplying all $g_i$ (or all $\ov{g}_i$) simultaneously results in a simple rescaling. Thus, effectively, it depends on $9$ coupling parameters. 

\subsection{Lax matrices}\label{lm} 
The Dunkl and Cherednik operators can be used to construct {\it Lax matrices} for the Calogero--Moser systems. We will first state the results in the case \eqref{elcm}.

In the setting of \ref{edo}, we have $\D(V)$ acting on $\c(V)$ by differential operators. Consider the induced module 
\begin{equation*}
M=\Ind_{\D(V)}^{\D(V)*W}\,\c(V)\,.
\end{equation*}
We write elements of $M$ as $f=\sum_{w\in W} w f_w$ with $f_w\in\c(V)$, thus identifying $M\cong\c W\otimes \c(V)$ as vector spaces. The algebra $\End_\c(M)$ is then identified with $\End_\c(\c W)\otimes \End_\c(\c(V))$, i.e. with \emph{operator-valued} matrices of size $|W|$. As a result, the action of $\D(V)*W$ on $M$ gives a representation
\begin{equation}\la{rep0}
\varrho\,:\ \D(V)*W\to \Mat(|W|, \D(V))\,.
\end{equation}
It is compatible with taking the classical limit, so we also get 
\begin{equation}\la{rep0c}
\varrho_c\,:\ A_0*W\to \Mat(|W|, A_0)\,,\qquad A_0=\c(V)\otimes \c[V^*]\,.
\end{equation}
Note that if $W'\subset W$ is a subgroup and $e'=\frac{1}{|W'|}\sum_{w\in W'} w$, then any $W'$-invariant element acts on $M$, preserving the subspace $M'=e'M$. Elements of $M'$ are $f=e'\sum_{u\in W'\backslash W} u f_u$, so \eqref{rep0}, \eqref{rep0c} restrict to
\begin{equation}\la{rep0'}
\varrho\,:\ (\D(V)*W)^{W'}\to \Mat(r, \D(V))\,, \qquad \varrho_{c}\,:\ (A_0*W)^{W'}\to \Mat(r, A_0)\quad (r=|W|/|W'|)\,.
\end{equation}
Recall that $A_0=\c(V)\otimes \c[V^*]$ is the classical limit of $\D(V)$, and it comes with the canonical Poisson bracket $\{\cdot, \cdot\}$. 

\begin{defi}
$(1)$ Given $\mathcal H\in\D(V)$ and $\mathcal L\in\Mat(r, \D(V))$, a {\it quantum Lax partner} of $\mathcal L$ relative to $\mathcal H$ is an element $\mathcal A\in \Mat(r, \D(V))$ such that $[\mathcal L, \mathcal H\mathbb I_r]=[\mathcal L, \mathcal A]$. That is, $[\mathcal L_{ij}, \mathcal H]=[\mathcal L, \mathcal A]_{ij}$ for $i,j=1,\dots, r$.

$(2)$ Given $H\in A_0$ and $L\in\Mat(r, A_0)$, a classical {\it Lax partner} of $L$ relative to $H$ is an element $A\in \Mat(r, A_0)$ such that $\{L_{ij}, H\}=[L, A]_{ij}$ for $i,j=1,\dots, r$. Equivalently, $\Dot{L}=[L,A]$ where $\Dot{L}$ denotes the time-derivative for the Hamiltonian flow defined by $H$. 
\end{defi}
\begin{remark}
The same definition applies with $\D(V)$ and $A_0$ replaced by the algebra of difference operators, $D_q(V)$, and its classical limit, $A_0=\c(V)[\Lambda]$, see \ref{clcaseq}.  \end{remark}
Now pick a weight $b\in P$ and consider Dunkl operators $y_\xi(\lambda)$ with $\lambda=z b$. The indeterminate $z\in \c$ will play the r\^ole of {\it spectral parameter} in the Lax matrix. Write $W'$ for the stabiliser of $b$, and $R'$ for the root system of $W'$. If $\lambda=zb$ then $\dpr{\alpha^\vee, \lambda}=0$ for $\alpha\in R'$, so $y_\xi(\lambda)$ may not be well defined due to the presence of terms $\dpr{\alpha, \xi}\sigma_{\dpr{\alpha^\vee, \lambda}}(\dpr{\alpha, x})$ with $\alpha\in R'$. However, if we specialise $\xi$ to $b$ then these terms drop out due to $\dpr{\alpha, \xi}=0$, giving a well-defined expression 
\begin{equation}\la{rdu}    
y_\xi(\lambda)=\hb\partial_\xi-\sum_{\alpha\in R_+\setminus R'_+}
k_\alpha \langle\alpha, \xi\rangle \sigma_{\langle\alpha^\vee, \lambda\rangle}(\langle\alpha, x\rangle)s_\alpha\,.
\end{equation}  
Its classical limit $y_{\xi, c}(\lambda)$ is obtained by replacing $\hb\partial_\xi$ with the classical momentum, $p_\xi$. 
Such $y_\xi(\lambda)$, $y_{\xi, c}(\lambda)$ are clearly $W'$-invariant.
Let $y, y_c$ denote these Dunkl operators (with $\xi=b$ and $\lambda=zb$). 
Define {\it quantum and classical Lax matrices} for the system \eqref{elcm} by
  \begin{equation}\label{lmat}
    \mathcal L=\varrho(y)\in \Mat(r, \D(V)),\qquad L=\varrho_c(y_c)\in \Mat(r, A_0) 
  \end{equation}
  in accordance with \eqref{rep0'}. These matrices depend on spectral parameter, $z$. 
  
  \begin{theorem} For $q\in\c[V^*]^W$, let $L_q\in \D(V)^W$ be the quantum Hamiltonian for the system \eqref{elcm}, constructed in Theorem \ref{efmv}, and $L_{q, c}\in A_0^W$ be its classical limit. Then,
for any $q\in\c[V^*]$, $\mathcal L$ (resp. $L$) has a quantum (resp. classical) Lax partner relative to $\mathcal H=L_q$ (resp. $H=L_{q, c}$).     
\end{theorem} 

\begin{cor}
 For any $b\in P$, let $r=|Wb|$. The classical elliptic Calogero--Moser system for a root system $R$ admits a Lax matrix $L$ of size $r$ with spectral parameter. Each of the commuting Hamiltonian flows of the system induces an isospectral deformation of $L$. The functions $h_k=\tr L^k$, $k\in\N$, form an involutive family, that is, $\{\tr\, L^a, \tr \,L^b\}=0$ for all $a,b\in\N$.      
\end{cor}

\bigskip

The difference analogues are dealt with in the same fashion. The setup is similar: in the setting of \ref{3.1}, we begin with the algebra of difference operators $\D_q(V)$ (associated with the lattice $\Lambda=P^\vee$) acting on the space of meromorphic functions, $\c(V)$. Replacing $\D(V)$ with $\D_q(V)$ in the above constructions, we consider the module 
\begin{equation*}
M=\Ind_{\D_q(V)}^{\D_q(V)*W}\,\c(V)\,.
\end{equation*}
The (left) action of $\D_q(V)*W$ on $M$ gives a representation
\begin{equation*}
\varrho\,:\ \D_q(V)*W\to \Mat(|W|, \D_q(V))\,.
\end{equation*}
It is compatible with taking the classical-limit map $\eta_0$, see \ref{clcaseq}, so we also get 
\begin{equation*}
\varrho_c\,:\ A_0*W\to \Mat(|W|, A_0)\,,\qquad A_0=\c(V)[P^\vee]\,.
\end{equation*}
\begin{lemma}
For a fundamental coweight $\lambda\in P^\vee_+$, write $W'$ for the stabiliser of $\lambda$ in $W$ and $e'$ for the corresponding symmetriser. Consider the elliptic Cherednik operator $Y:=Y^\lambda$, in which the spectral variable $\xi$ is specialised to $\xi=-\rho_k+z\lambda$, where $z\in\c$ and $\rho_k$ is given by \eqref{rho}. Then the action of the Cherednik operator $Y$ on $M$ preserves the subspace $M'=e'M$.      
\end{lemma}
Using the lemma, we introduce the {\it quantum Lax matrix} $\mathcal L$ by restricting 
$Y=Y^\lambda$ onto $M'$ and set the {\it classical Lax matrix} $L$ to be the classical limit of $\mathcal L$. The resulting matrices $\mathcal L\in \Mat(r, \D_q(V))$ and $L\in \Mat(r, A_0)$ are of size $r=|W|/|W'|$ and depend on spectral parameter, $z$. 

\begin{theorem} Let $L_b$ be one of the elliptic difference operators constructed in Theorem \ref{lb}, where $b\in P^\vee_+$ is (quasi-)minuscule. Write $L_{b, c}$ for the classical limit of $L_b$.
The Lax matrix $\mathcal L\in \Mat(r, \D_q(V))$ (resp. $L\in \Mat(r, A_0)$) has a quantum (resp. classical) Lax partner relative to $\mathcal H=L_b$ (resp. $H=L_{b, c}$). Hence, the Hamiltonian flow associated with $H=L_{b, c}$ induces an isospectral deformation of $L$.      
\end{theorem} 
This result is expected to remain valid for all commuting Hamiltonians $L_b$ constructed in Theorem \ref{emr}. This would imply that the classical Lax matrix remains isospectral under {\it all} commuting Hamiltonian flows. This is known to be true for $R=A_{n-1}$ and in the $\GL_n$-case (Ruijsenaars system), simply because all fundamental coweights are minuscule in that case. It has also been verified for the $C^\vee C_n$-case (van Diejen system).

\subsection{Historical comments} The elliptic Dunkl operators \eqref{edu} were introduced, and their commutativity proved, by Buchstaber, Felder and Veselov in \cite{BFV}. Etingof and Ma constructed Dunkl operators in a more general case, for any abelian variety $X$ with an action of a finite complex reflection group $W$ \cite{EM1}. They also considered and studied Cherednik algebra for such pairs $X,W$. The corresponding integrable systems made recent appearance in quantum filed theory (see \cite{ACL} for the detailed study of rank 1 case and further references). Theorem \ref{efmv} is proved in \cite{EFMV} for this more general case (including the Inozemtsev system \cite{I}). For the real crystallographic case \eqref{elcm}, another proof was given earlier by Cherednik in \cite{C4}.  

The constructions in Sections \ref{eqcase}--\ref{ecc} are due to Komori and Hikami \cite{KH98}, based on Cherednik's idea of Yang--Baxter equations on affine root systems \cite{C5}. Our account follows \cite{Ch} closely. The functional $R$-matrices \eqref{rije} go back to Shibukawa and Ueno \cite{SU}. The Hamiltonians \eqref{emr1}, \eqref{eru} were found (and showed to commute) by Ruijsenaars in \cite{R87}, see also \cite{RS} for the classical system and \cite{Ha} for a link to lattice models. For the $C^\vee C_n$-case (Van Diejen's system), van Diejen proposed in \cite{vD1} the Hamiltonian \eqref{clb1}--\eqref{clb2}  
under an additional constraint on the parameters. That constraint was removed and the integrability first proved in \cite{KH97} by a method, different to \cite{KH98}. Another, geometric proof was found recently by Rains within his geometric approach to elliptic DAHAs \cite{Rains}.  

Material in Section \ref{lm} is based on \cite{Ch}. The subject of Lax pairs for classical Calogero--Moser systems has a long history, see \cite{Ch} and references therein.  The classical matrices \eqref{lmat} reproduce those from \cite{DHP}, but their isospectrality under all commuting flows was not known prior to \cite{Ch}, except in type $R=A_{n-1}$. For the elliptic difference case with $R=A_{n-1}$, the Lax matrix was originally found by Ruijsenaars \cite{R87}, but nothing was known beyond $A_{n-1}$-case before \cite{Ch}, except for a partial result in the trigonometric case \cite{GP}. Notably, in \cite{Ch} the classical (as well as quantum) Lax matrix $L$ was found for the $C^\vee C_n$-case (van Diejen system), and its isospectrality under all commuting Hamiltonain flows was established.

\end{document}